\documentclass[11pt,a4paper]{article}

\usepackage{graphicx,epsf,epsfig,amssymb,amsmath,amsfonts,mathrsfs}
\usepackage{longtable}
\usepackage{bm}
\usepackage{color}
\usepackage{enumitem}
\usepackage{float}
\usepackage{caption}
\usepackage{graphicx}
\usepackage{booktabs}
\usepackage{array}
\usepackage{tabularx}
\usepackage{array}
\newcolumntype{Y}{>{\centering\arraybackslash}X}
\usepackage{enumitem}
\usepackage{caption}
\usepackage{subcaption}
\usepackage[section]{placeins}

\usepackage{subcaption}
\usepackage{upgreek}
\usepackage{cite}
\usepackage{mathtools}
\usepackage{titlesec}
\usepackage{lipsum}
\usepackage{xcolor}
\usepackage[section]{placeins}
\usepackage[colorlinks=true,linktocpage=true,linkcolor=blue,citecolor=blue]{hyperref}
\usepackage{tensor}
\usepackage{multirow}
\usepackage[utf8]{inputenc}
\usepackage{authblk}
\usepackage{array}

\usepackage{tikz}
\usepackage{booktabs}
\usepackage[normalem]{ulem}
\usepackage{soul}
\usepackage[toc,page]{appendix}

\newcommand{\be}{\begin{equation}}
\newcommand{\ee}{\end{equation}}
\newcommand{\bea}{\begin{eqnarray}}
\newcommand{\eea}{\end{eqnarray}}

\newcommand{\dmoff}[1]{}

\newcommand\blfootnote[1]{%
  \begingroup
  \renewcommand\thefootnote{}\footnote{#1}%
  \addtocounter{footnote}{-1}%
  \endgroup
}

\usepackage{comment}

\newcommand{\customrule}{%
\vspace{0.5cm}
\noindent\textcolor{gray}{\rule{\textwidth}{0.5pt}}
\vspace{0.5cm}
}

\usepackage[
a4paper,
left=1.5cm,
right=1.5cm,
top=3cm,
bottom=4cm,
]{geometry}

\numberwithin{equation}{section} 

\titleclass{\subsubsubsection}{straight}[\subsubsection]
\newcounter{subsubsubsection}[subsubsection]
\renewcommand\thesubsubsubsection{\thesubsubsection.\arabic{subsubsubsection}}

\titleformat{\subsubsubsection}[block]
  {\normalfont\normalsize\bfseries}
  {\thesubsubsubsection}{1em}{}

\titlespacing*{\subsubsubsection}
  {0pt}{3.25ex plus 1ex minus .2ex}{1em}

\makeatletter
\def\toclevel@subsubsubsection{4}
\def\l@subsubsubsection{\@dottedtocline{4}{7em}{4em}}
\makeatother

\author[]{Etevaldo dos Santos Costa Filho}

\affil[]{\normalsize Programa de Pós-Graduação em Física, Universidade Federal do Espírito Santo, Vitória, ES,  29075-910, Brazil}

\begin{document}
\title{\textbf{Electrically charged black holes with electroweak hair}}

\maketitle

\begin{abstract}
We construct electrically charged black holes in Einstein's gravity coupled to the bosonic sector of the electroweak theory, focusing on static, spherically symmetric and asymptotically flat solutions, with the weak mixing angle fixed to its Standard Model value. Horizon regularity imposes a resonance condition relating the charged-vector frequency to the electromagnetic potential at the horizon, corresponding to the threshold of charged superradiance. A perturbative analysis further shows that nonextremal Reissner-Nordstr\"om black holes do not support regular, exponentially localized, spherically symmetric electroweak clouds satisfying such condition. Thus, the hairy solutions cannot emerge via a bifurcation from this electrovacuum family. Moreover, our numerical investigation reveals two apparently disconnected families of bosonic stars. Within the explored parameter space, the hairy black holes are connected to the strongly gravitating one.
\end{abstract}
\vfill
\blfootnote{ \texttt{etevaldo.s.costa@ufes.br}}

	\newpage
	\customrule
	\tableofcontents

\customrule

\newpage

\section{Introduction}

Black holes have gone from purely exotic theoretical objects to an established part of our understanding of the Universe \cite{Herdeiro:2022yle}. The development of X-ray
astronomy and the identification of quasars provided an early observational context for the study of compact gravitational sources
\cite{Giacconi:1962zz,Schmidt:1963wkp}, while observations of stellar orbits provided further evidence for massive compact objects
at the Galactic centers \cite{Genzel:2010zy,Genzel:2024vou}. More recently, the detection of gravitational waves from binary black hole mergers
\cite{LIGOScientific:2016aoc} and the images obtained by the Event Horizon Telescope \cite{EventHorizonTelescope:2019dse,EventHorizonTelescope:2022wkp}
have opened complementary ways of probing these objects and the strong-field regime of gravity. These observations thus motivate the questions: are all these compact and massive objects, presumed to be black holes, indeed black holes? Could there exist other compact astrophysical objects capable of replicating part of the black hole phenomenology? Moreover, if they are black holes, are all of them described by the Kerr metric?

The Kerr hypothesis is the paradigm according to which the Kerr metric describes all astrophysical black holes in the Universe. This paradigm rests essentially on two main pillars. First, a set of mathematical uniqueness theorems showed that the most general stationary solution for physically relevant black holes, in vacuum General Relativity, is the Kerr solution. Second, inspired by these uniqueness theorems, a much more ambitious idea emerged, widely known as the ``no-hair conjecture'', proposed by John Wheeler and Remo Ruffini \cite{10.1063/1.3022513}. According to this idea, regardless of the initial matter/energy content, gravitational collapse leads to the formation of equilibrium black holes determined solely by their total mass, angular momentum (and possibly electric charge), all of which are asymptotically measured quantities governed by a Gauss law, with no other independent features (collectively called ``hair''). Under the astrophysically motivated assumption that black holes carry no electric charge, the no-hair conjecture would therefore imply the Kerr hypothesis: all astrophysical black holes in the Universe should be described by the Kerr metric, as it would be the only equilibrium solution capable of emerging dynamically.

On the other hand, over the past few decades, theoretical constructions of hairy black holes have become increasingly numerous in the literature as mechanisms to evade the no-hair theorems have been developed and understood. An early example was provided by Gibbons, who found black holes with scalar hair in an $\mathcal{N}=4$ supergravity model in 1982 \cite{Gibbons:1982ih}. Dilatonic and string-inspired charged black holes were subsequently studied in Refs.~\cite{Gibbons:1987ps,Garfinkle:1990qj}. The range of hairy configurations has since extended well beyond nonminimal couplings between scalar fields and electromagnetism \cite{Herdeiro:2025blx,Blazquez-Salcedo:2025cpu,Fernandes:2019kmh,dosSantosCostaFilho:2025ibq,Fernandes:2019rez,ChruscielCostaHeusler2012,Bizon:1994dh}. A different mechanism operates in rotating black holes with synchronized scalar or Proca hair, where hairy black holes exist at the threshold of superradiance instability  \cite{Herdeiro:2014goa,Herdeiro:2016tmi,Herdeiro:2024pmv,Brito:2015oca,East:2017ovw}.

It is worth emphasizing that many of these hairy black holes are constructed within simplified theories, often involving fields whose existence in Nature has not yet been established. Although such models provide useful laboratories for investigating strong gravity, a natural step towards more realistic configurations is to consider theories whose field content is known to exist in Nature. In this respect, non-Abelian gauge fields are particularly relevant, since they are an essential part of the Standard Model. Coupling these fields to gravity, therefore, provides a physically motivated setting for investigating hairy black holes and offers a natural opportunity for the high energy physics and gravity communities to work in synergy.

Historically, the inclusion of non-Abelian fields marked a turning point in the study of hairy black holes. It was somewhat surprising that the nonlinear structure of Yang-Mills theory could support black holes with nontrivial exterior fields despite the absence of corresponding static, regular, finite-energy solitons in
flat spacetime~\cite{VolkovGaltsov1989,KuenzleMasood1990,Bizon1990,
VolkovGaltsov1999,Deser:1976wq,Coleman:1977hd}. Their discovery was closely connected to that of globally regular, self-gravitating Yang-Mills configurations without horizons.

In 1988, Bartnik and McKinnon found a discrete family of globally regular, topologically trivial solutions in four-dimensional Einstein-Yang-Mills (EYM) theory with gauge group $\mathrm{SU}(2)$ \cite{Bartnik:1988am}. The existence of these self-gravitating configurations relies on gravity, since pure Yang-Mills theory admits no corresponding static, finite-energy solitons in flat spacetime \cite{Deser:1976wq,Coleman:1977hd}. Their black hole extensions were found soon afterwards \cite{VolkovGaltsov1989,KuenzleMasood1990,Bizon1990}, providing explicit counterexamples to the generalized no-hair conjecture. These static, asymptotically flat black holes support nontrivial Yang-Mills fields despite having vanishing global Yang-Mills charges. The families are labeled by the number of nodes of the gauge-field amplitude, and, for a fixed node number, their zero-horizon-radius limit recovers the corresponding Bartnik-McKinnon soliton \cite{VolkovGaltsov1999}. The numerical constructions were subsequently supplemented by existence proofs \cite{BreitenlohnerForgacsMaison1994}. Similar families of globally regular self-gravitating solutions and their black hole generalization have also been constructed in Yang-Mills-Higgs theories, including gravitating monopoles and their black hole generalizations \cite{Greene:1992fw,Klinkhamer1984,Kleihaus1991,Kunz1992,Breitenlohner:1991aa,Lee:1991jw,Herdeiro:2026pzt,Brihaye:2004nd,Jain:2022kwq,PhysRevD.110.104045}.

Since these discoveries, more than three decades of work have established many properties of non-Abelian black holes. Static EYM black holes need not be spherically symmetric \cite{KleihausKunz1997,RidgwayWeinberg1995}. Perturbative analyses also show that a black hole horizon can rotate even when the total angular momentum vanishes~\cite{BrodbeckEtAl1997}, while fully rotating non-Abelian solutions have been constructed as well \cite{KleihausKunz2001}. The corresponding staticity theorems involve additional conditions on the electric sector \cite{PhysRevD.46.1453,PhysRevD.47.R5209}. Moreover, circularity does not follow automatically for stationary, axisymmetric Yang-Mills fields; thus, the use of a Lewis-Papapetrou metric requires further assumptions \cite{Heusler:1993cj,ChruscielCostaHeusler2012}. The standard sigma-model structure underlying electrovacuum uniqueness arguments is also not directly reproduced by the Yang-Mills system \cite{VolkovGaltsov1999,ChruscielCostaHeusler2012}.  Within the Einstein-Weinberg-Salam model, magnetically charged black holes supporting nontrivial electroweak fields outside the horizon have been constructed~\cite{Bai2021,Gervalle:2024yxj,Gervalle2025}. These solutions provide a natural motivation for investigating whether electroweak hair can also be supported by purely electrically charged black holes.

In the previous work \cite{CostaFilhoGervalle2026}, we constructed static, spherically symmetric and asymptotically flat bosonic stars in the Einstein-Weinberg-Salam model, with the electroweak couplings  fixed to their Standard Model values, but at a different overall energy scale\footnote{Flat-spacetime solutions were studied in Ref.~\cite{EWballs}. Within the families investigated, solutions at the measured Standard Model couplings and mass ratios were not found.}. In the present work, we investigate whether such field configurations may admit a regular event horizon while retaining nontrivial exterior fields. Therefore, we construct and study the hairy black hole generalization, focusing on static, spherically symmetric, purely electrically charged solutions with the weak mixing angle fixed to its Standard Model value, $\sin^2\theta_{\mathrm{W}}=0.223$ \cite{PhysRevD.110.030001}.

The rest of the paper is organized as follows. Section~\ref{sec:model} introduces the Einstein-Weinberg-Salam model, its field equations and residual gauge freedom, and derives the horizon regularity conditions, the resonance condition, a mass formula and the corresponding first law. Section~\ref{sec:spherical-ansatz} specifies the static, spherically symmetric ansatz and the boundary conditions. Section~\ref{sec:numerical-results} presents the numerical results for the bosonic stars and hairy black holes, while Sec.~\ref{sec:remarks} discusses our main conclusions. Appendix~\ref{app:eom} displays the equations of motion specialized to the spherical ansatz. Appendix~\ref{app:analytic-obstructions} derives integral identities constraining the electric sector, and Appendix~\ref{app:rn-zero-mode} examines the absence of regular, localized, spherically symmetric charged-vector clouds at linear order on nonextremal Reissner-Nordstr\"om backgrounds.

\section{The Einstein-Weinberg-Salam model}
\label{sec:model}

\subsection{Action, field equations and energy-momentum tensor}
\label{sec:model-equations}

We adopt the dimensionless formulation of the bosonic Einstein-Weinberg-Salam
model \cite{Weinberg:1967tq,Salam:1968rm}, used in Refs.~\cite{CostaFilhoGervalle2026,Gervalle2025,Gervalle:2024yxj}.
The matter fields are a complex Higgs doublet $\Phi=(\phi_1,\phi_2)^T$,
an $\mathrm{SU}(2)$ gauge potential $W=T_aW^a_\mu\,\mathrm{d}x^\mu$, and
the $\mathrm{U}(1)$ hypercharge potential $B=B_\mu\,\mathrm{d}x^\mu$.
Here $T_a=\tau_a/2$ are the $\mathrm{SU}(2)$ generators. These fields are
minimally coupled to Einstein's gravity, and we use the metric signature
$(-,+,+,+)$. The action reads
\begin{equation}
 S=\frac{e^2}{4\pi\alpha}\int\left(
 \frac{R}{2\kappa}+\mathcal{L}_{\mathrm{WS}}
 \right)\sqrt{-\mathrm{g}}\,\mathrm{d}^4x,
 \label{eq:dimensionless-action}
\end{equation}
where $\kappa$ and $\beta$ are the dimensionless gravitational and Higgs self-couplings, respectively, and $\alpha$ is the electromagnetic fine-structure constant. The bosonic electroweak Lagrangian reads
\begin{align}
\mathcal{L}_{\mathrm{WS}}=
-\frac{1}{4g^{2}}\,W^{a}_{\mu\nu}W^{a\mu\nu}
-\frac{1}{4g'^{2}}\,B_{\mu\nu}B^{\mu\nu}
-(D_{\mu}\Phi)^{\dagger}D^{\mu}\Phi
-\frac{\beta}{8}\left(\Phi^{\dagger}\Phi-1\right)^{2}.
\nonumber
\end{align}

The corresponding field strengths and the gauge-covariant derivative
of the Higgs doublet are defined by
\begin{align}
 W^a_{\mu\nu}
 =\partial_\mu W^a_\nu-\partial_\nu W^a_\mu
 +\epsilon_{abc}W^b_\mu W^c_\nu,\quad
 B_{\mu\nu}
 =\partial_\mu B_\nu-\partial_\nu B_\mu,\quad
 D_\mu\Phi
 =\left(\partial_\mu-\mathrm{i}T_aW^a_\mu
 -\frac{\mathrm{i}}{2}B_\mu\right)\Phi.
 \label{eq:Higgs-covariant-derivative}
\end{align}

For the analysis below, we adopt the 't Hooft definitions of the
electromagnetic and $Z$ fields \cite{tHooft:1974kcl,Nambu:1977ag,Gervalle2025}. We work in the unitary gauge,
$\Phi=(0,\phi)^T$, with real Higgs amplitude $\phi$ . We combine the first two $\mathrm{SU}(2)$ components
into a complex charged vector field. This field and its field strength
are defined by
\begin{equation}
    w_\mu=\frac{1}{g}(W^1_\mu+\mathrm{i}W^2_\mu),
    \qquad
    w_{\mu\nu}=\mathcal{D}_\mu w_\nu-\mathcal{D}_\nu w_\mu,
    \label{eq:w-proca-field}
\end{equation}
with $\mathcal{D}_\mu=\nabla_\mu+\mathrm{i}W^3_\mu$. In this gauge,
the electromagnetic and $Z$ potentials are given by
\begin{equation}
    A_\mu=\frac{g}{g'}B_\mu+\frac{g'}{g}W^3_\mu,
    \qquad
    Z_\mu=B_\mu-W^3_\mu.
\end{equation}

These definitions and their relation to the alternative definitions
due to Nambu are discussed in
Refs.~\cite{tHooft:1974kcl,Nambu:1977ag,Gervalle2025}. With the coupling normalization $e=gg'$ and $g^2+g'^2=1$, the
covariant derivative takes the form
\begin{equation*}
    \mathcal{D}_\mu
    =\nabla_\mu+\mathrm{i}(eA_\mu-g^2Z_\mu).
\end{equation*}

We also introduce the neutral field strengths and the condensate tensor
\begin{align*}
    F_{\mu\nu}&=\nabla_\mu A_\nu-\nabla_\nu A_\mu,
    \qquad
    Z_{\mu\nu}=\nabla_\mu Z_\nu-\nabla_\nu Z_\mu,\\
    \psi_{\mu\nu}&=\frac{\mathrm{i}}{2}
    (w_\mu\bar{w}_\nu-w_\nu\bar{w}_\mu).
\end{align*}

The tensor $\psi_{\mu\nu}$ is real and antisymmetric. In terms of
these variables, the Lagrangian can be written as
\begin{align}
 \mathcal{L}_{\mathrm{WS}}={}&
 -\frac{1}{4}F_{\mu\nu}F^{\mu\nu}
 -\frac{1}{4}Z_{\mu\nu}Z^{\mu\nu}
 -\frac{e}{2}F_{\mu\nu}\psi^{\mu\nu}
 +\frac{g^2}{2}Z_{\mu\nu}\psi^{\mu\nu}
 -\frac{1}{4}w_{\mu\nu}\bar{w}^{\mu\nu}
 \notag\\
 &-\frac{g^2}{4}\psi_{\mu\nu}\psi^{\mu\nu}
 -\partial_\mu\phi\,\partial^\mu\phi
 -\frac{1}{4}\left(g^2w_\mu\bar{w}^\mu+Z_\mu Z^\mu\right)\phi^2
 -\frac{\beta}{8}(\phi^2-1)^2 .
 \label{eq:unitary-lagrangian}
\end{align}

Varying the action with respect to the matter fields $(A_\nu,Z_\nu,w_\nu,\phi)$ yields
\begin{align}
 \nabla_\mu F^{\nu\mu}&=eJ^\nu,
 \label{eq:maxwell-equation}\\
 \nabla_\mu Z^{\nu\mu}+\frac{1}{2}\phi^2Z^\nu&=-g^2J^\nu,
 \label{eq:Z-equation}\\
 \mathcal{D}_\mu w^{\mu\nu}
 +\mathrm{i}\left(g^2\psi^{\nu\sigma}+eF^{\nu\sigma}
 -g^2Z^{\nu\sigma}\right)w_\sigma
 &=\frac{g^2}{2}\phi^2w^\nu,
 \label{eq:w-equation}\\
 \nabla_\mu\nabla^\mu\phi
 -\frac{1}{4}\left(g^2w_\mu\bar{w}^\mu+Z_\mu Z^\mu\right)\phi
 &=\frac{\beta}{4}(\phi^2-1)\phi.
 \label{eq:Higgs-equation}
\end{align}

The current appearing in the equations for $A_\mu$ and $Z_\mu$ is
\begin{equation}
 J^\nu=\nabla_\mu\psi^{\mu\nu}
 +\frac{\mathrm{i}}{2}\left(
 w_\mu\bar{w}^{\mu\nu}-\bar{w}_\mu w^{\mu\nu}
 \right).
 \label{eq:electric-current}
\end{equation}

Its conservation follows from the antisymmetry of $F^{\mu\nu}$ in
Eq.~\eqref{eq:maxwell-equation}:
\begin{equation}
 \nabla_\nu J^\nu=0.
 \label{eq:current-conservation}
\end{equation}

Taking the gauge-covariant divergence of Eq.~\eqref{eq:w-equation}
and the covariant divergence of Eq.~\eqref{eq:Z-equation} gives,
respectively
\begin{align}
 \mathcal{D}^\mu(\phi^2w_\mu)&=-\mathrm{i}\phi^2Z^\mu w_\mu,
 \label{eq:w-constraint}\\
 \nabla_\mu(\phi^2Z^\mu)&=0.
 \label{eq:Z-constraint}
\end{align}

These Lorenz-like conditions follow from the vector equations and do not
provide additional independent evolution equations. The variation of the action with respect to the metric leads to Einstein's equations
\begin{equation}
 R_{\mu\nu}-\frac{1}{2}\mathrm{g}_{\mu\nu}R=\kappa T_{\mu\nu},
 \label{eq:Einstein-equations}
\end{equation}
with the energy-momentum tensor
\begin{align}
    T_{\mu\nu}=&\; F_{\mu\sigma}F_\nu{}^\sigma + Z_{\mu\sigma}Z_\nu{}^\sigma + e\left(F_{\mu\sigma}\psi_\nu{}^\sigma+F_{\nu\sigma}\psi_\mu{}^\sigma\right)-g^2\left(Z_{\mu\sigma}\psi_\nu{}^\sigma+Z_{\nu\sigma}\psi_\mu{}^\sigma\right)+g^2\psi_{\mu\sigma}\psi_\nu{}^\sigma\notag\\
    &+\frac{1}{2}\left(\bar{w}_{\mu\sigma}w_\nu{}^\sigma+\bar{w}_{\nu\sigma}w_\mu{}^\sigma\right)+2\partial_\mu\phi\,\partial_\nu\phi+\frac{1}{4}g^2\phi^2\left(\bar{w}_\mu w_\nu+\bar{w}_\nu w_\mu\right)+\frac{1}{2}\phi^2 Z_\mu Z_\nu+\mathrm{g}_{\mu\nu}\mathcal{L}_{\mathrm{WS}}.
\end{align}

Finally, the original  $\mathrm{SU}(2)\times\mathrm{U}(1)$ gauge potentials can be reconstructed by inverting the field definitions
\begin{equation}
 \begin{aligned}
 W={}&\left[
 \frac{g}{2}(w_\mu+\bar{w}_\mu)T_1
 -\frac{\mathrm{i}g}{2}(w_\mu-\bar{w}_\mu)T_2
 +(eA_\mu-g^2Z_\mu)T_3
 \right]\mathrm{d}x^\mu,\\
 B={}&(eA_\mu+g'^2Z_\mu)\,\mathrm{d}x^\mu.
 \end{aligned}
 \label{eq:original-gauge-potentials}
\end{equation}

\subsection{Residual gauge freedom}
\label{sec:gauge-inheritance}

The unitary gauge leaves a residual electromagnetic gauge freedom associated with $\mathrm{U}(1)_{\mathrm{em}}$. For a real gauge parameter $\lambda(x)$, the remaining local transformations are of the form
\begin{align}\label{eq:residual-gauge}
    \mathrm{U}(1)_{\mathrm{em}}\; :\quad w_\mu\to e^{-\mathrm{i}\lambda(x)}w_\mu,\quad \bar{w}_\mu\to e^{\mathrm{i}\lambda(x)}\bar{w}_\mu,\quad A_\mu\to A_\mu+\frac{1}{e}\nabla_\mu\lambda(x),\quad Z_\mu\to Z_\mu,\quad \phi\to\phi.
\end{align}

Thus, $A_\mu$ is the electromagnetic connection. The neutral massive
vector $Z_\mu$ and the Higgs amplitude $\phi$ remain unchanged under these
transformations, as are $F_{\mu\nu}$, $Z_{\mu\nu}$ and $\psi_{\mu\nu}$.
The charged vector $w_\mu$ and its field strength $w_{\mu\nu}$ both
transform with the same local phase.

In this work, we focus on asymptotically flat, spherically symmetric, static black holes in four dimensions.
Let $\xi^\mu$ denote the static Killing field of the asymptotically flat
configurations considered here. Since $Z_\mu$ and $\phi$ are real, they inherit the spacetime symmetry
\begin{equation}
 \mathcal{L}_\xi Z_\mu=0,
 \qquad
 \mathcal{L}_\xi\phi=0.
 \label{eq:neutral-inheritance}
\end{equation}

In the charged sector, the fields are invariant up to a residual electromagnetic gauge transformation \cite{Bergmann:1978fi,Forgacs:1979zs}. Introducing a real function $\chi$, we write
\begin{equation}
 \mathcal{L}_\xi A_\mu=\frac{1}{e}\nabla_\mu\chi,
 \qquad
 \mathcal{L}_\xi w_\mu=-\mathrm{i}\chi w_\mu.
 \label{eq:charged-inheritance}
\end{equation}

Under a residual electromagnetic gauge transformation with parameter $\lambda$, the function $\chi$ transforms according to
\begin{equation}
 \chi\to\chi+\mathcal{L}_\xi\lambda.
 \label{eq:chi-transformation}
\end{equation}

We therefore introduce the gauge-invariant combination
\begin{equation}
 X\equiv\frac{\chi}{e}-\xi^\mu A_\mu.
 \label{eq:X-definition}
\end{equation}

The symmetry conditions above then yield the covariant identities
\begin{align}
 \nabla_\mu(\xi^\nu Z_\nu)&=\xi^\nu Z_{\mu\nu},
 \label{eq:Z-Killing-identity}\\
 \nabla_\mu X&=\xi^\nu F_{\nu\mu},
 \label{eq:X-Killing-identity}\\
 \mathcal{D}_\mu(\xi^\nu w_\nu)&=\xi^\nu w_{\mu\nu}
 -\mathrm{i}\left[eX+g^2\xi^\nu Z_\nu\right]w_\mu.
 \label{eq:w-Killing-identity}
\end{align}

The antisymmetry of $F_{\mu\nu}$ in Eq.~\eqref{eq:X-Killing-identity}
also gives $\mathcal{L}_\xi X=0$. These identities will be used in the
horizon analysis and the derivation of the mass formula in what follows.

\subsection{Horizon properties and the resonance condition}
\label{sec:covariant-horizon}

We now derive the horizon conditions, using the symmetry assumptions of
Sec.~\ref{sec:gauge-inheritance}, required for the existence of such hairy black holes. The starting point is the vanishing of
the null contraction of the energy-momentum tensor on a Killing horizon.
Related geometric arguments can be found, for instance, in
\cite{Smolic:2016dmh,Herdeiro:2024pmv,Prabhu:2015vua,Herdeiro:2025blx}.

Assume that the spacetime possesses a regular, connected Killing horizon
$\mathcal{H}$ generated by the static Killing field $\xi^\mu$, normalized
by $\xi^\mu\xi_\mu\to-1$ at spatial infinity\footnote{Although our focus in this paper lies on static black holes, the generalization of the horizon properties to the stationary case is straightforward.}. The vector $\xi^\mu$ is
timelike throughout the static exterior and null on $\mathcal{H}$.
We assume that the matter fields are regular in a local unitary gauge and
that the Higgs amplitude satisfies $\phi|_{\mathcal{H}}\neq0$.  One can
show that the contraction of the Ricci tensor with the horizon generator vanishes on $\mathcal{H}$ \cite{Bardeen:1973gs,heusler_1996}
\begin{equation}
 R_{\mu\nu}\xi^\mu\xi^\nu\big|_{\mathcal{H}}=0.
 \label{eq:horizon-Ricci-identity}
\end{equation}

Using Einstein's equations and the fact that $\xi$ becomes null on the horizon, this condition translates into a constraint on the matter content at the horizon
\begin{align}
 0={}&\left.T_{\mu\nu}\xi^\mu\xi^\nu\right|_{\mathcal{H}} =\Bigg(
 \left[\xi^\alpha(F_{\alpha\mu}+e\psi_{\alpha\mu})\right]
 \left[\xi^\beta(F_\beta{}^\mu+e\psi_\beta{}^\mu)\right] +\left[\xi^\alpha(Z_{\alpha\mu}-g^2\psi_{\alpha\mu})\right]
 \left[\xi^\beta(Z_\beta{}^\mu-g^2\psi_\beta{}^\mu)\right]
 \notag\\
 &+(\xi^\alpha\bar{w}_{\alpha\mu})(\xi^\beta w_\beta{}^\mu)
 +2(\xi^\mu\nabla_\mu\phi)^2
 +\frac{g^2}{2}\phi^2|\xi^\mu w_\mu|^2
 +\frac{1}{2}\phi^2(\xi^\mu Z_\mu)^2
 \Bigg)\Bigg|_{\mathcal{H}}.
 \label{eq:horizon-square-decomposition}
\end{align}

The scalar derivative term in
Eq.~\eqref{eq:horizon-square-decomposition} vanishes identically due to the symmetry inheritance \eqref{eq:neutral-inheritance}. For any antisymmetric tensor $H_{\mu\nu}$, the vector
$\xi^\nu H_{\nu\mu}$ is orthogonal to $\xi^\mu$.  Its norm on a null
hypersurface is nonnegative and can vanish only when it is proportional to
the null generator. Thus every term in
Eq.~\eqref{eq:horizon-square-decomposition} is nonnegative and must vanish
separately.  Consequently, for every vector $s^\mu$ tangent to $\mathcal{H}$
\begin{align}
 s^\mu\xi^\nu(F_{\nu\mu}+e\psi_{\nu\mu})\big|_{\mathcal{H}}= s^\mu\xi^\nu(Z_{\nu\mu}-g^2\psi_{\nu\mu})\big|_{\mathcal{H}}=  s^\mu\xi^\nu w_{\nu\mu}\big|_{\mathcal{H}}=0,
 \label{eq:horizon-tangential-strengths}
\end{align}
\begin{equation}
 \xi^\mu w_\mu\big|_{\mathcal{H}}= \xi^\mu Z_\mu\big|_{\mathcal{H}}=0.
 \label{eq:horizon-mass-contractions}
\end{equation}

The definition of $\psi_{\mu\nu}$ then gives
\begin{equation}
 \xi^\nu\psi_{\nu\mu}\big|_{\mathcal{H}}
 =\left.\frac{\mathrm{i}}{2}\left[
 (\xi^\nu w_\nu)\bar{w}_\mu
 -w_\mu(\xi^\nu\bar{w}_\nu)\right]\right|_{\mathcal{H}}=0.
 \label{eq:horizon-psi-contraction}
\end{equation}

Consequently
\begin{equation}
 s^\mu\xi^\nu F_{\nu\mu}\big|_{\mathcal{H}}= s^\mu\xi^\nu Z_{\nu\mu}\big|_{\mathcal{H}}=0.
 \label{eq:horizon-individual-strengths}
\end{equation}

Equation~\eqref{eq:X-Killing-identity} now gives
$s^\mu\nabla_\mu X|_{\mathcal{H}}=0$, implying
\begin{equation}
 X\big|_{\mathcal{H}}=X_H=\text{constant}.
 \label{eq:horizon-X-constant}
\end{equation}

This is the constancy of the horizon potential associated with the
combined spacetime and electromagnetic symmetry. It is consistent with
the general treatment of horizon potentials in theories with internal
gauge freedom \cite{Prabhu:2015vua}. We next contract Eq.~\eqref{eq:w-Killing-identity} with a vector $s^\mu$ tangent to $\mathcal{H}$.  Since $\xi^\nu w_\nu$ vanishes everywhere on $\mathcal{H}$, its gauge-covariant derivative in every tangential direction vanishes.
Together with Eq.~\eqref{eq:horizon-tangential-strengths}, this gives
\begin{equation}
eX_H\,s^\mu w_\mu\big|_{\mathcal{H}}=0.
 \label{eq:horizon-tangential-resonance}
\end{equation}

Equation~\eqref{eq:horizon-tangential-resonance} gives $X_H=0$ whenever
$w_\mu$ has a nonvanishing tangential component on $\mathcal{H}$. For the
complementary case, assume that the nonextremal horizon is a bifurcate Killing horizon. We further assume
that, for a nontrivial charged configuration, there exists a finite $p\geq0$ such that
$\mathcal{W}^{(p)}\neq0$ on the bifurcation surface $\mathcal{B}$, where
$\mathcal{W}^{(p)}_{\alpha_1\ldots\alpha_p\mu}
\equiv\mathcal{D}_{\alpha_1}\cdots\mathcal{D}_{\alpha_p}w_\mu$. For
clarity, consider first the case
$\mathcal{W}^{(1)}_{\alpha\mu}\equiv\mathcal{D}_\alpha w_\mu\neq0$.

 On the bifurcation surface $\mathcal{B}$, $\xi^\mu=0$ and
$\nabla_\mu\xi_\nu=\kappa_H(m_\mu n_\nu-n_\mu m_\nu)$, where $n^\mu$ and $m^\mu$ form an orthonormal pair normal to $\mathcal{B}$, with $n^\mu$ timelike and $m^\mu$ spacelike \cite{Gourgoulhon2026,Iyer:1994ys}. The matrix $K^\mu{}_\nu\equiv\nabla_\nu\xi^\mu$ is
diagonalizable with eigenvalues $-\kappa_H,+\kappa_H,0,0$,  where $\kappa_H$ is the surface gravity defined in Eq.~\eqref{eq:surface-gravity-definition}. The Lie derivative,
$\mathcal{L}_\xi\mathcal{W}^{(1)}=-\mathrm{i}\chi\mathcal{W}^{(1)}$, can then be written on the bifurcation surface
\begin{equation}
 K^\rho{}_\alpha\mathcal{W}^{(1)}_{\rho\mu}
 +K^\rho{}_\mu\mathcal{W}^{(1)}_{\alpha\rho}
 =-\mathrm{i} eX_H\mathcal{W}^{(1)}_{\alpha\mu}
 \qquad\text{on }\mathcal{B} .
 \label{eq:first-derivative-bifurcation}
\end{equation}

Choose eigenvectors $\vartheta^{(a)}_\mu$ satisfying
$K^\rho{}_\mu\vartheta^{(a)}_\rho
=\lambda_a\vartheta^{(a)}_\mu$, with
$\lambda_a\in\{-\kappa_H,+\kappa_H,0,0\}$. Expanding
$\mathcal{W}^{(1)}$ in this basis
\begin{equation}
 \mathcal{W}^{(1)}_{\alpha\mu}
 =\sum_{a,b}C_{ab}\vartheta^{(a)}_\alpha\vartheta^{(b)}_\mu,
\end{equation}
Eq.~\eqref{eq:first-derivative-bifurcation} gives
\begin{equation}
 \bigl(\lambda_a+\lambda_b+\mathrm{i} eX_H\bigr)C_{ab}=0 .
\end{equation}

Because $\mathcal{W}^{(1)}\neq0$, some $C_{ab}$ is nonzero, and hence
$-\mathrm{i} eX_H=\lambda_a+\lambda_b$. The right-hand side is real whereas the
left-hand side is purely imaginary. Hence
\begin{equation}\label{eq:horizon-resonance-condition}
    X_H=0.
\end{equation}

This is the so-called ``resonance condition'' \cite{HerdeiroRadu2020}.
The same argument applies for every finite $p\geq0$. Under the assumptions
stated above, this is therefore a necessary condition for the existence
of black holes with nontrivial electroweak hair. 


\subsection{A mass formula}
\label{sec:mass-formula}

In an asymptotically flat, static spacetime, let $\Sigma$ be an exterior spacelike hypersurface bounded by a horizon cross-section $\mathcal{S}_H$
and a two-sphere $\mathcal{S}^2_\infty$ at spatial infinity. The Komar mass $\mathcal{M}$ associated with $\xi^\mu$ is then given by the  integral
\begin{equation}
 \mathcal{M}=-\frac{1}{\kappa}\int_{\mathcal{S}^2_\infty}
 \nabla^\mu\xi^\nu\,\mathrm{d} S_{\mu\nu}.
 \label{eq:BH-Komar-mass}
\end{equation}

Using the Killing equation then gives \cite{Bardeen:1973gs}
\begin{equation}
 \mathcal{M}=-\frac{1}{\kappa}\int_{\mathcal{S}_H}
 \nabla^\mu\xi^\nu\,\mathrm{d} S_{\mu\nu}
 -\frac{2}{\kappa}\int_\Sigma R^\mu{}_{\nu}\xi^\nu
 \,\mathrm{d}\Sigma_\mu.
 \label{eq:BH-Komar-Ricci-decomposition}
\end{equation}

By means of Einstein's equations, we have
\begin{equation}
 \mathcal{M}=\frac{2\kappa_H A_H}{\kappa}
 -2\int_\Sigma\left(
 T^\mu{}_{\nu}-\frac{1}{2}\delta^\mu_\nu T
 \right)\xi^\nu\,\mathrm{d}\Sigma_\mu ,
 \label{eq:BH-Komar-decomposition}
\end{equation}
where $A_H$ is the black hole's area and $\kappa_H$ is the surface gravity defined by
\begin{equation}
 \xi^\nu\nabla_\nu\xi^\mu\big|_{\mathcal{H}}
 =\kappa_H\xi^\mu\big|_{\mathcal{H}},
 \qquad
 \kappa_H^2=-\frac{1}{2}(\nabla_\mu\xi_\nu)
 (\nabla^\mu\xi^\nu)\big|_{\mathcal{H}}.
 \label{eq:surface-gravity-definition}
\end{equation}

Combining Eqs.~\eqref{eq:maxwell-equation}--\eqref{eq:Higgs-equation} and
\eqref{eq:Z-Killing-identity}--\eqref{eq:w-Killing-identity}, we can write on shell \cite{CostaFilhoGervalle2026}
\begin{equation}
 \left(T^\mu{}_{\nu}-\frac{1}{2}\delta^\mu_\nu T\right)\xi^\nu
 =\nabla_\sigma\mathcal{V}^{\mu\sigma}
 +\left(\frac{R}{2\kappa}+\mathcal{L}_{\mathrm{WS}}\right)\xi^\mu,
 \label{eq:Komar-integrand-reduction}
\end{equation}
where
\begin{align}
 \mathcal{V}_{\mu\nu}={}
 (\xi^\rho Z_\rho)(-Z_{\mu\nu}+g^2\psi_{\mu\nu})
 -\frac{1}{2}\left[
 \bar{w}_{\mu\nu}(\xi^\rho w_\rho)
 +w_{\mu\nu}(\xi^\rho\bar{w}_\rho)
 \right]+(F_{\mu\nu}+e\psi_{\mu\nu})X .
 \label{eq:V-tensor}
\end{align}

Defining the on-shell Lagrangian by
\begin{equation}
 L\equiv\int_\Sigma\left(
 \frac{R}{2\kappa}+\mathcal{L}_{\mathrm{WS}}
 \right)\xi^\mu\,\mathrm{d}\Sigma_\mu .
 \label{eq:spatial-L}
\end{equation}

Substitution into Eq.~\eqref{eq:BH-Komar-decomposition} gives
\begin{equation}
 \mathcal{M}=\frac{2\kappa_HA_H}{\kappa}
 -2\int_\Sigma\nabla_\sigma\mathcal{V}^{\mu\sigma}
 \,\mathrm{d}\Sigma_\mu-2L.
 \label{eq:mass-before-V-boundaries}
\end{equation}

Moreover, the residual symmetry \eqref{eq:residual-gauge} leaves the Lagrangian
invariant, giving rise to the on-shell conserved current
\begin{equation*}
 j_\lambda^\mu= \nabla_\nu(F^{\mu\nu}+e\psi^{\mu\nu}),
 \qquad \nabla_\mu j_\lambda^\mu=0.
\end{equation*}

Hence, the electric charge can be defined as a Komar integral and, therefore, being evaluated as the flux over a two-sphere at infinity. Due to the exponential decay of the $w$-boson, the quantity $\psi_{\mu\nu}$ does not contribute and the electric charge is defined as
\begin{equation}
 Q_e=-\frac{1}{8\pi}\int_{\mathcal{S}^2_\infty}
 F^{\mu\nu}\,\mathrm{d} S_{\mu\nu}.
 \label{eq:surface-charges}
\end{equation}

We are now able to compute the explicit contribution of the quantity $\mathcal{V}$ to the mass formula. Using Stokes' theorem, we might write
\begin{equation}
    \int_\Sigma\nabla_\sigma\mathcal{V}^{\mu\sigma}
 \,\mathrm{d}\Sigma_\mu=\frac{1}{2}\int_{\mathcal{S}^2_\infty-\mathcal{S}_H}\mathcal{V}^{\mu\nu}\,\mathrm{d} S_{\mu\nu}=-4\pi X_\infty Q_e,
\end{equation}
where we have used the fact that the massive fields decay exponentially at infinity. On the horizon, we assume
$\phi|_{\mathcal{H}}\neq0$, so that the positive terms in Eq.~\eqref{eq:horizon-square-decomposition} give the
contractions~\eqref{eq:horizon-mass-contractions}. The horizon value of $X$ is zero by Eq.~\eqref{eq:horizon-resonance-condition}. Consequently, the black hole mass formula is
\begin{equation}
 \mathcal{M}=\frac{2\kappa_HA_H}{\kappa}
 +8\pi X_\infty Q_e-2L .
 \label{eq:covariant-BH-mass-formula}
\end{equation}

Notice that the mass formula \eqref{eq:covariant-BH-mass-formula} is gauge invariant. We can further work out the above formula by rewriting the Lagrangian density as
\begin{align}
 \left(\frac{R}{2\kappa}+\mathcal{L}_{\mathrm{WS}}\right)\xi^\mu
 ={}-\frac{1}{2}\nabla_\nu\mathcal{V}^{\mu\nu}
 +\frac{\beta}{8}(\phi^2-1)^2\xi^\mu
 +\frac{\phi^2}{4}Z^\mu(\xi^\nu Z_\nu)+\frac{g^2\phi^2}{8}\left[
 w^\mu(\xi^\nu\bar{w}_\nu)
 +\bar{w}^\mu(\xi^\nu w_\nu)\right].
 \label{eq:on-shell-L-density}
\end{align}

Therefore, after integrating the Lagrangian term, the mass formula becomes
\begin{align}
 \mathcal{M}={}&\frac{2\kappa_H A_H}{\kappa}
 +4\pi X_\infty Q_e
 -2\int_\Sigma\Bigg\{
 \frac{\beta}{8}(\phi^2-1)^2\xi^\mu
 +\frac{\phi^2}{4}Z^\mu(\xi^\nu Z_\nu)
 \notag\\
 &\hspace{29mm}
 +\frac{g^2\phi^2}{8}\left[
 w^\mu(\xi^\nu\bar{w}_\nu)
 +\bar{w}^\mu(\xi^\nu w_\nu)\right]\Bigg\}\,\mathrm{d}\Sigma_\mu.
 \label{eq:covariant-electric-BH-mass-formula}
\end{align}

Following the approaches in \cite{heusler_1996,Heusler:1993cj}, we can derive the  first law
\begin{equation}
    \delta\mathcal{M}=(\kappa_H/\kappa)\,\delta A_H+4\pi X_\infty\,\delta Q_e.
\end{equation}

Finally, in the gauge used in the present paper, $A_t(\infty)=0$, and hence $X_\infty=\omega/e$, as we shall discuss in Sec.~\ref{sec:spherical-ansatz}. 

\section{Spherically symmetric ansatz}
\label{sec:spherical-ansatz}

We now specialize to static, spherically symmetric, asymptotically flat configurations and fix the gauge freedom. We assume that all fields are regular on the horizon and throughout the exterior region.  We consider the following line element
\begin{equation}
 \mathrm{d} s^2=-e^{2F_0(r)}N(r)\,\mathrm{d} t^2
 +e^{2F_1(r)}\left[
 \frac{\mathrm{d} r^2}{N(r)}+r^2\left(
 \mathrm{d}\theta^2+\sin^2\theta\,\mathrm{d}\varphi^2
 \right)\right],
 \qquad
 N(r)=1-\frac{r_H}{r}.
 \label{eq:BH-metric-ansatz}
\end{equation}

For $r_H>0$, $r$ is not the isotropic or
areal radius in general; it is a Schwarzschild-like coordinate adapted to the
horizon.  Setting $r_H=0$ gives $N=1$, and the metric ansatz takes the
isotropic-coordinate form used for the horizonless solutions of
Ref.~\cite{CostaFilhoGervalle2026}. Assuming regularity at $r=r_H$, this hypersurface is a Killing horizon generated by the static Killing field $\xi^\mu=(\partial_t)^\mu$.  If $F_0$ and $F_1$ are finite at $r_H$, its area and surface gravity are
\begin{equation}
 A_H=4\pi r_H^2e^{2F_1(r_H)},
 \qquad
 \kappa_H =\frac{e^{F_0(r_H)-F_1(r_H)}}{2r_H}.
 \label{eq:horizon-area-gravity}
\end{equation}

In turn, we take the matter ansatz of the form
\begin{equation}
 A=A_t(r)\,\mathrm{d}t ,
 \qquad
 Z=Z_t(r)\,\mathrm{d}t,
 \qquad
 w=e^{-\mathrm{i}\omega t}
 \bigl(w_t(r)\,\mathrm{d}t+\mathrm{i} w_r(r)\,\mathrm{d}r\bigr),
 \qquad
 \phi=\phi(r),
 \label{eq:BH-matter-ansatz}
\end{equation}
with real radial profiles. Hence, the seven unknown functions are
\begin{equation}
 \{F_0,F_1,\phi,w_t,w_r,Z_t,A_t\}.
 \label{eq:BH-radial-functions}
\end{equation}

Substituting the spherical ansatz~\eqref{eq:BH-matter-ansatz}
into Eq.~\eqref{eq:original-gauge-potentials}, we obtain
\begin{equation}
 \begin{aligned}
 W={}&\Big[
 g w_t(r)\big(T_1\cos(\omega t)-T_2\sin(\omega t)\big)
 +\big(eA_t(r)-g^2Z_t(r)\big)T_3
 \Big]\mathrm{d}t\\
 &+g w_r(r)\big(T_1\sin(\omega t)+T_2\cos(\omega t)\big)
 \,\mathrm{d}r,\\
 B={}&\big(eA_t(r)+g'^2Z_t(r)\big)\,\mathrm{d}t.
 \end{aligned}
 \label{eq:spherical-gauge-potentials}
\end{equation}

We emphasize that, although the fields $w$ and, consequently, $W$ depend explicitly on time, the resulting energy-momentum tensor is time independent.

For the numerical construction, it is convenient to parametrize the exterior region by a radial coordinate that vanishes at the horizon. We introduce a new radial coordinate $x$ via
\begin{equation}
 x=\sqrt{r^2-r_H^2},
 \label{eq:horizon-coordinate}
\end{equation}
so that the event horizon is located at $x= 0$ and spatial infinity corresponds to $x\to\infty$.  Assuming that the unknown functions are analytic in $r$ at $r=r_H$, when rewritten in terms of the coordinate $x$, these become even functions of $x$. Hence, all the unknowns~\eqref{eq:BH-radial-functions} have vanishing first derivatives with respect to $x$ at the horizon, $x=0$. In practice, we impose these Neumann conditions on $F_0$, $F_1$ and $\phi$, while $w_t$, $w_r$, $Z_t$ and $A_t$ take the Dirichlet values given in Eqs.~\eqref{eq:basic-matter-horizon-BC} and \eqref{eq:horizon-evenness-conditions} below.

At spatial infinity, asymptotic flatness and finite energy require the fields to approach the vacuum. Linearizing the field equations in the asymptotic region yields the large-$x$ behavior
\begin{equation}
\begin{aligned}
F_0(x)&=-\frac{M-r_H/2}{x}+\mathcal{O}(x^{-2}),
&\qquad F_1(x)&=\frac{M-r_H/2}{x}+\mathcal{O}(x^{-2}),\\
A_t(x)&=A_t(\infty)+\frac{Q_e}{x}+\mathcal{O}(x^{-2}),
& Z_t(x)&=c_Z\frac{e^{-m_{\mathrm{Z}}x}}{x}+\cdots,\\
\phi(x)&=1+c_H\frac{e^{-m_{\mathrm{H}}x}}{x}+\cdots,
& w_t(x)&=c_W\frac{e^{-kx}}{x}+\cdots,\\
w_r(x)&=\frac{c_W\left[\omega-eA_t(\infty)\right]}{k}\,
\frac{e^{-kx}}{x}+\cdots.
\end{aligned}
\end{equation}

Here $M,Q_e,c_Z,c_H,c_W$ are real coefficients, and we have introduced the decay constant
\begin{equation}
    k=\sqrt{m_{\mathrm{W}}^2-[\omega-eA_t(\infty)]^2}.
\end{equation}

The vacuum masses are $m_{\mathrm{W}}=g/\sqrt{2}$, $m_{\mathrm{Z}}=1/\sqrt{2}$, and $m_{\mathrm{H}}=\sqrt{\beta/2}$. Hence,  any localized solution satisfies the bound-state condition $|\omega-eA_t(\infty)|<m_{\mathrm{W}}$.  The $1/x$ term in $A_t$ is the Coulomb tail of the massless electromagnetic field and, therefore, $Q_e$ is the total electric charge of the configuration, including the horizon flux and the charge carried by the exterior fields. Similarly, $M$ is the ADM mass, related to the Komar mass by $M=\kappa\mathcal{M}/(8\pi)$.  

Having established the gauge-invariant  resonance condition \eqref{eq:horizon-resonance-condition}, we may evaluate it in the harmonic time-dependent gauge we adopt for the charged massive  $w$-boson. Choosing $\chi=\omega$, there remains a residual gauge freedom of the form $\lambda=\lambda_0+ct$, with $\lambda_0$ and $c$ constants. The time-dependent part of this freedom is fixed by imposing $A_t(\infty)=0$, while $\lambda_0$ only induces an irrelevant constant phase rotation of $w_\mu$\footnote{Notice that, in fact, one can choose a gauge such that all fields are static, but when doing so, one loses the freedom to impose $A_t(\infty)=0$ \cite{Bizon:1992pi,VolkovGaltsov1999}.}\textsuperscript{,\,}{}\footnote{The field equations also constrain the sign of the electric charge. The positive identity derived in Appendix~\ref{app:positive-identity} implies $\omega Q_e>0$.}. The  condition \eqref{eq:horizon-resonance-condition} therefore reads
\begin{equation}
 \omega=eA_t\big|_{\mathcal{H}}.
 \label{eq:horizon-harmonic-resonance}
\end{equation}

Analogous electric resonance conditions arise for nonlinear scalar and electroweak $Q$-clouds \cite{HongSuzukiYamada2020,HerdeiroRaduEWClouds2026} and for fully backreacting black holes with charged scalar hair \cite{HerdeiroRadu2020,Kunz:2023qfg}. Moreover, it is well known that, in pure $\mathrm{SU}(2)$ Einstein-Yang-Mills theory, static, spherically symmetric and asymptotically flat configurations with nonzero charge cannot be globally regular, while black holes  possessing a regular event horizon are necessarily embedded Abelian Reissner-Nordstr\"om \cite{GaltsovErshov1989,ErshovGaltsov1990,Bizon:1992pi}. This restriction does not extend generally to theories containing a Higgs field \cite{Lee:1991jw,VolkovGaltsov1999,Greene:1992fw}. Hence, the Einstein-Weinberg-Salam model evades such theorems.

\bigskip

Finally, for the spherical ansatz, the horizon regularity and resonance conditions translate into the following boundary conditions:
\begin{equation}
 w_t(0)=0,
 \qquad
 Z_t(0)=0,
 \qquad
 A_t(0)=\frac{\omega}{e}.
 \label{eq:basic-matter-horizon-BC}
\end{equation}

In the numerical scheme, we supplement these conditions with
\begin{equation}
 \partial_xF_0=\partial_xF_1=\partial_x\phi=w_r=0,
 \label{eq:horizon-evenness-conditions}
\end{equation}
while at spatial infinity, we impose
\begin{equation}
 F_0=F_1=A_t=Z_t=w_t=w_r=0,\qquad \phi=1.
 \label{eq:infinity}
\end{equation}


\section{Numerical results}\label{sec:numerical-results}

The numerical solutions are specified by five independent input parameters, namely $\theta_{\mathrm{W}}$, $\kappa$, $\beta$, $\omega$ and $r_H$. The weak mixing angle $\theta_{\mathrm{W}}$ fixes the relative strength of the two gauge couplings through $g=\cos\theta_{\mathrm{W}}$ and $g'=\sin\theta_{\mathrm{W}}$. The parameter $\kappa$ controls the strength of the gravitational coupling, while $\beta$ fixes the Higgs self-coupling. The frequency $\omega$ determines the harmonic time dependence of the charged vector field $w_\mu$. Finally, $r_H$ fixes the position of the horizon. We construct a family of solutions fixing
\begin{equation}
    \sin^2 \theta_{\mathrm{W}}=0.223\,,\qquad \kappa=0.98\,,\qquad \beta=0.02\,
\end{equation}
and varying the parameters $\omega$ and $r_H$. That is, we have systematically explored the black hole existence domain by varying the horizon radius, $r_H$, for constant angular frequencies, $\omega$, and vice versa. In total, we have obtained thousands of black hole solutions, along sequences of constant frequency, $0.586\leq\omega\leq0.598$ (that is, $0.940\lesssim\omega/m_{\mathrm{W}}\lesssim0.959$), and sequences of constant horizon radius, $2.26\times10^{-4}\leq r_H\leq0.07$. The choice $\beta=0.02$ yields
$m_{\mathrm{H}}/m_{\mathrm{W}}=\sqrt{\beta}/g\simeq0.16$, which is below the physical value 
$m_{\mathrm{H}}/m_{\mathrm{W}}\simeq1.56$ corresponds to
$\beta\simeq1.88$~\cite{Gervalle2025,PhysRevD.110.030001}.
The numerical construction became progressively more challenging as
$\beta$ was increased.

The stars and black hole solutions were computed using two independent numerical schemes, based on the finite-element code FreeFEM~\cite{MR3043640} and the finite-difference code CADSOL~\cite{schmauder1992cadsol,SCHONAUER1989279} (see also Ref.~\cite{BSFactory} for a comparison between the codes). For all solutions we have found, the metric and matter functions, together with their first and second derivatives, have smooth profiles. This leads to finite curvature invariants on the black hole's exterior region, in particular at the event horizon. As a test of the numerical accuracy, we have evaluated, for every black hole, the errors measured by Lorenz-like constraint \eqref{eq:w-constraint}, the one unsolved component of Einstein's equations and the relative difference between the Komar mass integral and the ADM mass. The overall errors remain below $10^{-8}$. The resonance condition~\eqref{eq:horizon-harmonic-resonance} is imposed as a boundary condition.

\subsection{The two families of bosonic stars}

\begin{figure}[htbp]
    \centering
    \begin{subfigure}[b]{0.485\linewidth}
        \centering
        \includegraphics[width=\linewidth]{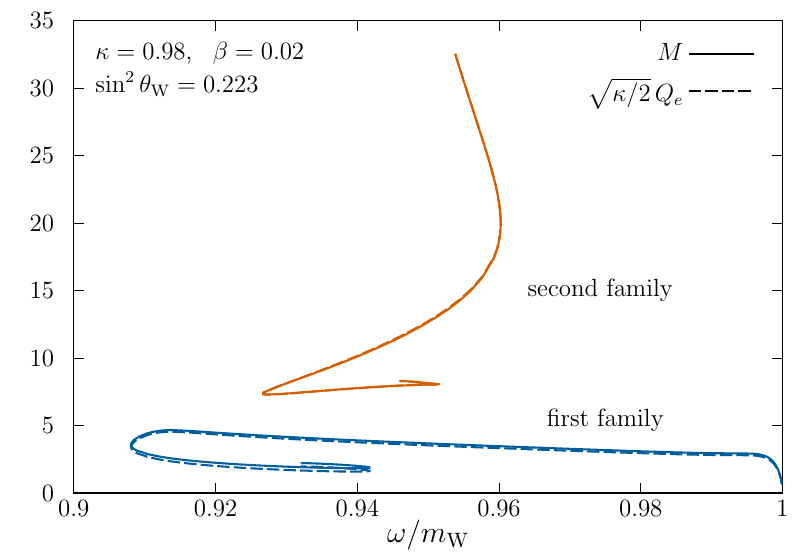}
    \end{subfigure}\hfill
    \begin{subfigure}[b]{0.485\linewidth}
        \centering
        \includegraphics[width=\linewidth]{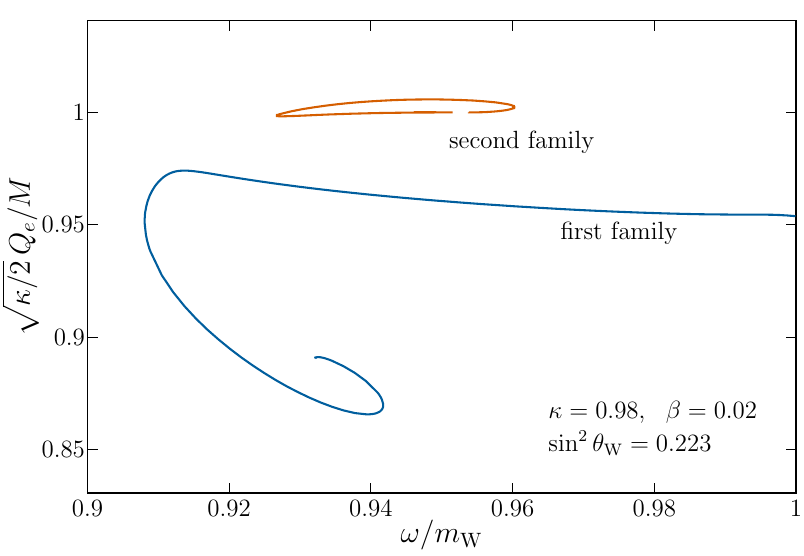}
    \end{subfigure}
    \caption{(Left panel) The ADM mass $M$ (solid) and the rescaled electric charge $\sqrt{\kappa/2}\,Q_e$ (dashed) as functions of $\omega/m_{\mathrm{W}}$ for the two families of bosonic stars, with $\kappa=0.98$, $\beta=0.02$ and $\sin^2\theta_{\mathrm{W}}=0.223$. For the second family the two curves almost coincide within the line width. (Right panel) The corresponding ratio $\sqrt{\kappa/2}Q_e/M$ as a function of $\omega/m_{\mathrm{W}}$. }
    \label{fig:massstar}
\end{figure}

When constructing hairy black holes with fields allowing for horizonless self-gravitating solutions, a natural question that arises is whether such families are connected. That is, starting from a hairy black hole, is it possible to continuously shrink the horizon up to the limit where it vanishes and the end state is a horizonless self-gravitating condensate? When trying to answer such a question, we found (apparently) two disconnected families of bosonic stars where only one of them was reported in \cite{CostaFilhoGervalle2026} (see Fig.~\ref{fig:massstar}).

The first is the family reported in Ref.~\cite{CostaFilhoGervalle2026}, which connects continuously to dilute configurations. Its mass-frequency diagram exhibits the characteristic spiral structure familiar from bosonic stars~\cite{Liebling:2012fv,Brito:2015pxa,herdeiro2023procahiggs}. The dilute solutions correspond to the Newtonian limit, approached as $\omega \to m_{\mathrm{W}}$. With the model parameters held fixed, decreasing $\omega$ away from this limit yields increasingly compact configurations whose mass initially increases, reaches a maximum, and then decreases. The frequency continues to decrease until it reaches $\omega_\text{min}$, where the sequence backbends and develops successive branches of the spiral.

For the second family, neither a weak-field limit nor a connection to the first family was found within the explored solutions. These configurations remain strongly gravitating throughout the  whole parameter space, although gravitational effects are less pronounced in its intermediate region and become stronger towards both ends. Increasing numerical errors prevented further continuation in either direction, leaving the ultimate extent of the family undetermined. The hairy black holes constructed here are connected to this second family of stars.

The second family does in fact produce a stronger gravity regime. The compactness of such stars, as defined in \cite{CostaFilhoGervalle2026}, lives in the range $0.55\lesssim\mathcal{C}\lesssim 1$, whereas $0\leq\mathcal{C}\lesssim0.32$ for the first family. For comparison, an extremal RN black hole has $\mathcal{C}=1$ . Moreover, light rings are found only on the fourth branch of the first family, while they are present throughout the explored parameter space of the second family.

The heavier end of the second family deserves some comments. The most massive computed solution on the upper part of the curve in Fig.~\ref{fig:massstar} has $\omega/m_{\mathrm W}\simeq0.954$ and $M\simeq32.53$ ($Mm_{\mathrm W}\simeq20.27$), with $\sqrt{\kappa/2}\,Q_e/M-1\simeq2.5\times10^{-9}$ and $R_{99}\simeq M$. Its central potential nearly satisfies the relation expected at a resonant black hole horizon, $1-eA_t(0)/\omega\simeq10^{-4}$. The frequency is close to the critical value
\begin{equation}
 \omega_c\equiv\frac{e}{\sqrt{\kappa/2}}\,,\qquad \frac{\omega_c}{m_{\mathrm{W}}}\simeq0.954\,,
 \label{eq:critical-frequency}
\end{equation}
which is the frequency for which the resonance condition is satisfied on the horizon of the extremal RN black hole. Indeed, Eq.~\eqref{eq:app-RN-synchronization} gives $\omega=eQ_e/R_+\leq\omega_c$, since $R_+\geq M\geq\sqrt{\kappa/2}\,Q_e$, the bound being saturated only in the extremal limit, $R_+=M=\sqrt{\kappa/2}\,Q_e$. These features could suggest that, at this end, the stars approach the extremal RN black hole, the matter fields being confined inside an increasingly deep gravitational well, as in the quasi-black hole limit \cite{Lemos:2007yh}. The numerical errors, however, grow in this limit, and we could not follow the family further.

\subsection{Hairy black holes}

We now present the numerical results for the constructed hairy black holes. We present the domain of existence in terms of the dimensionless quantities $\omega/m_{\mathrm{W}}$, $Mm_{\mathrm{W}}$ and $\sqrt{\kappa/2}\,Q_e/M$. Part of the boundary of the domain of existence is provided by the second family of stars (solid line), which corresponds to the limit $r_H\to 0$, while the remaining part of the boundary (dashed lines) is given by the last computed solutions along our sequences, and does not necessarily represent a physical boundary.

\begin{figure}[htbp]
    \centering
    \begin{subfigure}[b]{0.485\linewidth}
        \centering
        \includegraphics[width=\linewidth]{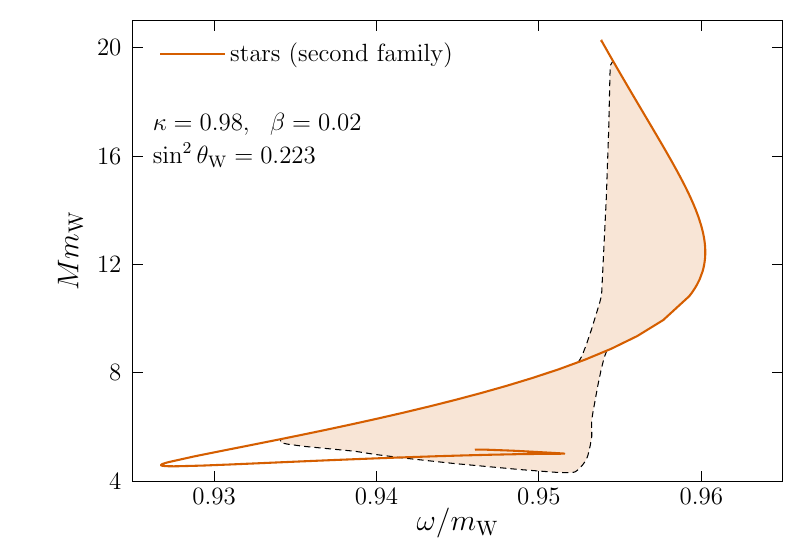}
    \end{subfigure}\hfill
    \begin{subfigure}[b]{0.485\linewidth}
        \centering
        \includegraphics[width=\linewidth]{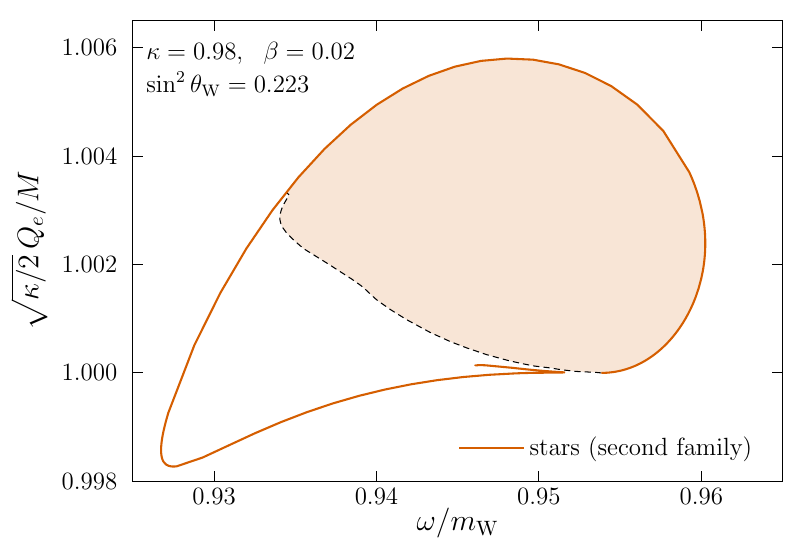}
    \end{subfigure}
    \caption{Domain of existence of the hairy black holes (shaded region), for $\kappa=0.98$, $\beta=0.02$ and $\sin^2\theta_{\mathrm{W}}=0.223$. (Left panel) The ADM mass $Mm_{\mathrm{W}}$ as a function of $\omega/m_{\mathrm{W}}$. (Right panel) The ratio $\sqrt{\kappa/2}\,Q_e/M$ as a function of $\omega/m_{\mathrm{W}}$. The solid line corresponds to the second family of bosonic stars (see Fig.~\ref{fig:massstar}), which is the $r_H\to0$ limit of the black holes. The dashed lines are numerical curves of the last obtained black hole solutions.}
    \label{fig:domain}
\end{figure}

The structure of the domain of existence can be understood, for instance, by following sequences with constant frequency. Each sequence of black hole solutions emerges from a star in the middle part of the family, that is, between the maximal and the minimal frequencies in Fig.~\ref{fig:domain}. Along a first branch of stars, starting from the maximal mass and decreasing up to the backbending around $\omega/m_{\mathrm{W}}\approx0.960$,  the horizon radius increases from zero up to a maximal value, after which it decreases again, connecting the first and second star branches up to a frequency  very close to the critical frequency $\omega_c$, defined in \eqref{eq:critical-frequency}. These black holes fill the upper part of the domain in the left panel of Fig.~\ref{fig:domain}, between the two parts of the star curve.

In the stars second branch, for $\omega<\omega_c$, on the other hand, the black holes fill the lower part of the domain.  They, likewise, emerge from the stars in the second branch, with $r_H$ increasing from zero to a maximum before decreasing again. However, within the explored solutions, these sequences do not return to the third or subsequent branches of the star family. The Hawking temperature decreases along the returning black hole branch, suggesting that they are approaching an extremal black hole. Increasing numerical errors prevented further continuation, so this limiting behavior could not be established. The right panel of Fig.~\ref{fig:domain} shows that all the black holes violate the RN bound $\sqrt{\kappa/2}\,Q_e\leq M$, with $1.000007\leq\sqrt{\kappa/2}\,Q_e/M\leq1.0058$. This is not surprising, since the stars to which they are connected also violate such a limit.

\begin{figure}[htbp]
    \centering
    \includegraphics[width=0.485\linewidth]{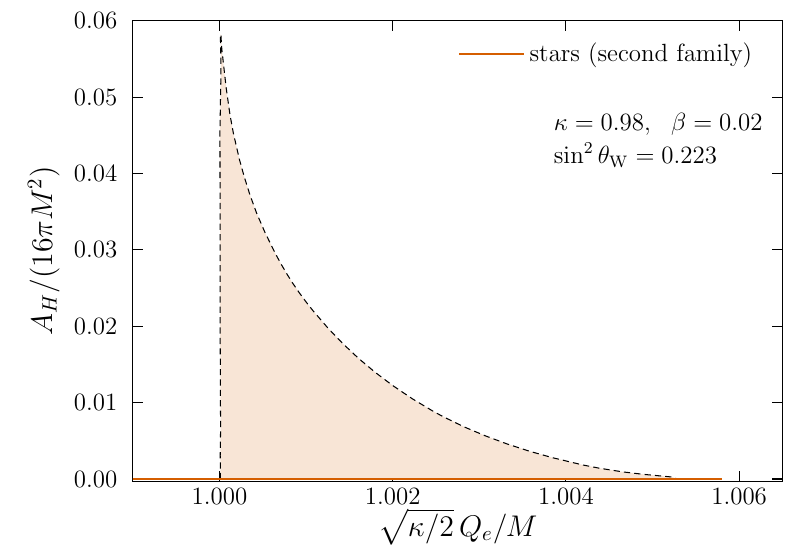}
    \caption{The reduced horizon area $A_H/(16\pi M^2)$ as a function of $\sqrt{\kappa/2}\,Q_e/M$ for the hairy black holes (shaded region), with $\kappa=0.98$, $\beta=0.02$ and $\sin^2\theta_{\mathrm{W}}=0.223$.  (solid line). The dashed lines are numerical curves of the last obtained solutions.}
    \label{fig:area}
\end{figure}

The reduced horizon area, $A_H/(16\pi M^2)$, is shown in Fig.~\ref{fig:area} as a function of $\sqrt{\kappa/2}\,Q_e/M$. The reduced area is normalized such that it is 1 for the Schwarzschild black hole. The stars, having no horizon, lie on the line with vanishing area. All computed hairy black holes have small reduced horizon areas, while RN black holes have reduced areas between 0.25 and 1, the lower value being attained in the extremal limit. The largest values of the reduced area are found close to the RN bound, whereas the black holes with the largest charge-to-mass ratio, $\sqrt{\kappa/2}\,Q_e/M>1.005$, have very small reduced horizon areas.

It is also useful to consider a possible connection with electrovacuum black holes. Appendix~\ref{app:rn-zero-mode} shows that a nonextremal
RN black hole does not support a regular, exponentially localized, spherically symmetric charged-vector zero mode satisfying
the resonance condition. Consequently, within the sector and regularity assumptions considered here, the hairy black holes cannot emerge through
a perturbative bifurcation from the RN family. This is consistent with our numerical results: since all the hairy black holes violate the RN bound, none of them has the same mass and electric charge as a RN black hole. The two families only come close to each other in the neighborhood of $\omega_c$, where the extremal RN black hole would be the natural limiting configuration. However, the black holes with the smallest charge-to-mass ratio we have found, $\sqrt{\kappa/2}\,Q_e/M-1\simeq7\times10^{-6}$, still have a reduced area of approximately $0.044$, far from the extremal RN value $1/4$, although their temperature is already small, $8\pi T_H M\simeq8\times10^{-3}$. Thus, we cannot decide whether such a limit is actually attained.

\begin{figure}[htbp]
    \centering
    \begin{subfigure}[b]{0.485\linewidth}
        \centering
        \includegraphics[width=\linewidth]{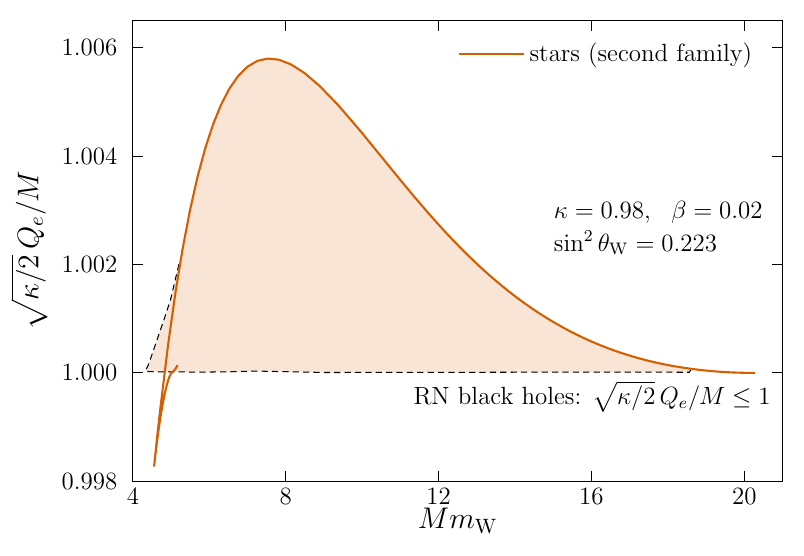}
    \end{subfigure}\hfill
    \begin{subfigure}[b]{0.485\linewidth}
        \centering
        \includegraphics[width=\linewidth]{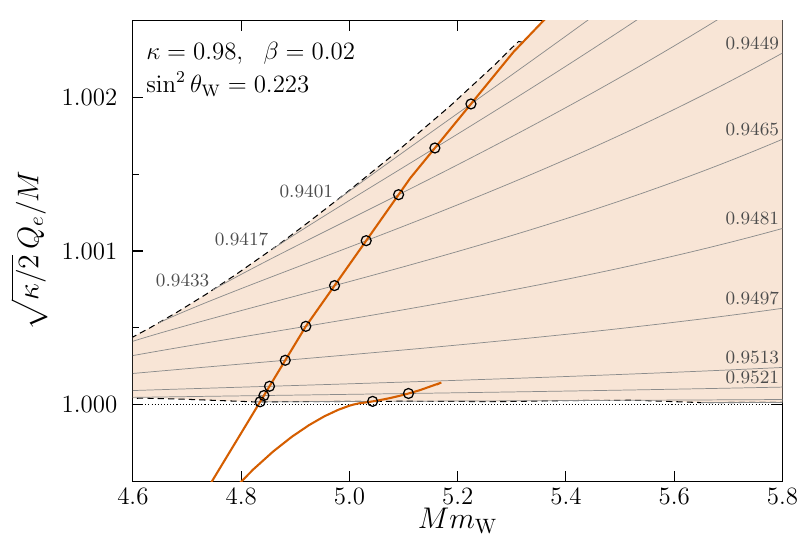}
    \end{subfigure}
    \caption{(Left panel) Domain of existence of the hairy black holes (shaded region) in a $\sqrt{\kappa/2}\,Q_e/M$ vs. $Mm_{\mathrm{W}}$ diagram. The solid line corresponds to the second family of stars and the dashed lines are numerical curves of the last obtained solutions. RN black holes have $\sqrt{\kappa/2}\,Q_e/M\leq1$. (Right panel) Zoom of the region $4.6\leq Mm_{\mathrm{W}}\leq5.8$. The grey lines are sequences of black holes with constant frequency, labeled by $\omega/m_{\mathrm{W}}$. Each circle marks a star and a black hole with the same mass and electric charge. The dotted line is the RN bound $\sqrt{\kappa/2}\,Q_e=M$.}
    \label{fig:nonuniqueness}
\end{figure}

An interesting feature of the computed solution space is that bosonic stars and hairy black holes can share the same mass and electric charge. Thus, the global charges $(M,Q_e)$ do not uniquely specify a configuration within the combined stars and black hole solution space. To illustrate this, we display the numerically explored domain in terms of the global charges in Fig.~\ref{fig:nonuniqueness} (left panel), where $\sqrt{\kappa/2},Q_e/M$ is shown as a function of $Mm_{\mathrm{W}}$.

In Fig.~\ref{fig:nonuniqueness} (right panel), we show a zoom of this region. The grey lines are sequences of black holes with constant frequency, labeled by $\omega/m_{\mathrm{W}}$, and the solid line corresponds to the stars. As one can see, these sequences cross the star curve. That is, each circle in the figure corresponds to two different configurations, a star and a hairy black hole, with the same global charges $(M,Q_e)$. A natural question is whether two hairy black holes can also share the same global charges. We found no evidence for such case in the explored parameter space.

For the graphical representation of the field profiles, it is convenient to use an areal radius $R=r e^{F_1}$ and introduce a compactified radial coordinate $s$ via
\begin{equation}
 s=\frac{R}{R+3R_H}.
\end{equation}

\begin{figure}[htbp]
    \centering
    \begin{subfigure}[b]{0.485\linewidth}
        \centering
        \includegraphics[width=\linewidth]{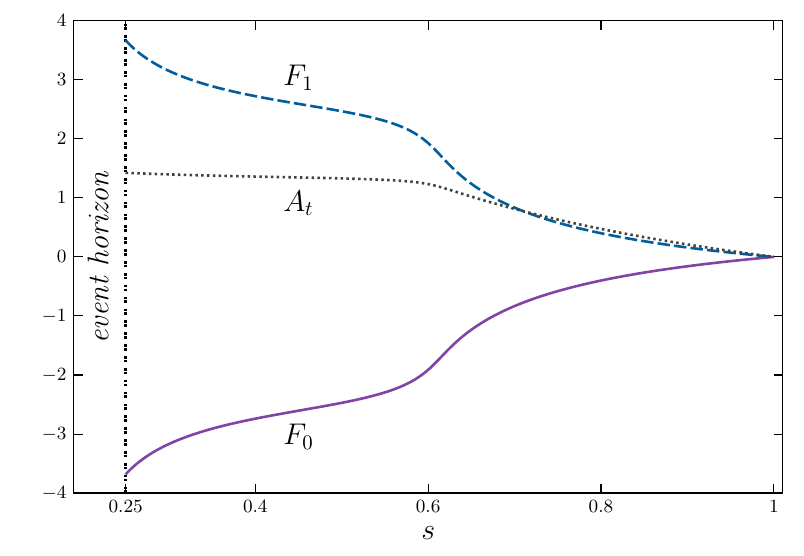}
    \end{subfigure}\hfill
    \begin{subfigure}[b]{0.485\linewidth}
        \centering
        \includegraphics[width=\linewidth]{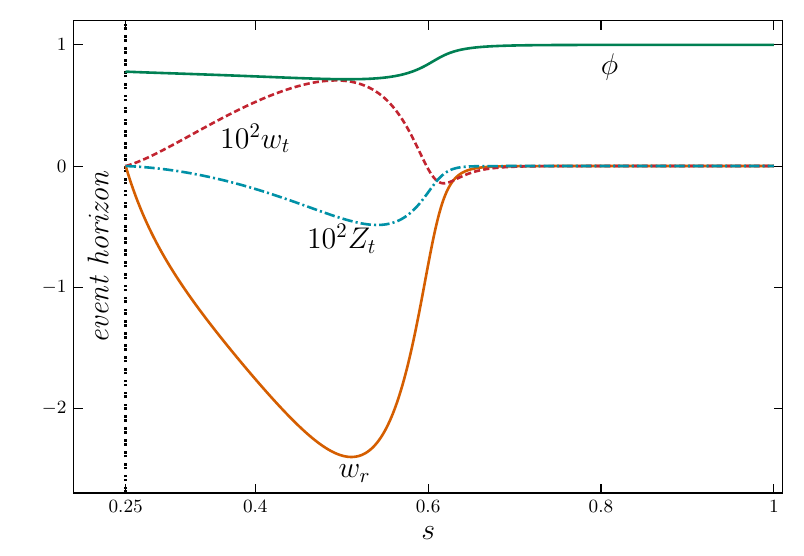}
    \end{subfigure}
    \caption{Profile functions of a typical solution in the second branch with $\omega=0.593$ and $r_H=0.078$, versus the compactified radial coordinate $s$. Left: $A_t$, $F_0$ and $F_1$. Right: $Z_t$, $w_t$, $w_r$ and $\phi$. The functions $Z_t$ and $w_t$ are multiplied by $10^2$ for visibility. The vertical dotted line mark the event horizon at $s=1/4$.}
    \label{fig:radial_profiles}
\end{figure}

The event horizon is located at $s=1/4$, while spatial infinity corresponds to $s\to1$. Thus, the complete exterior region is represented within $1/4\leq s<1$.  The typical profiles of $F_0$, $F_1$, $A_t$, $Z_t$, $w_t$, $w_r$ and $\phi$ are shown in Fig.~\ref{fig:radial_profiles} versus the compactified coordinate $s$, for a solution in the second branch. The metric functions $F_0$ and $F_1$ are smooth, remain finite at the horizon and approach zero at infinity. The electromagnetic potential $A_t$ decreases from $A_t(r_H)=\omega/e$ to zero, whereas $w_t$, $w_r$ and $Z_t$ vanish at both boundaries. Notice that $w_t$ changes sign. For the displayed solution, the Higgs field $\phi$  reaches a minimum outside the horizon and approaches its vacuum value $\phi=1$ at infinity.

This solution belongs to the second branch of the sequence with $\omega/m_{\mathrm{W}}\simeq0.9513$, close to its maximal horizon radius, $r_H\simeq0.0782$. It has $Mm_{\mathrm{W}}\simeq7.55$, $\sqrt{\kappa/2}\,Q_e/M\simeq1.0015$, $a_H\simeq0.016$ and $t_H\simeq0.20$, its horizon areal radius being $R_H\simeq0.25M$. 

To display the spherical profiles in a plane, we further define
\begin{equation}
    X=s\cos\varphi,\qquad Y=s\sin\varphi, \qquad 0\leq\varphi<2\pi.
\end{equation}

 It follows that $X^2+Y^2=s^2$. The horizon is therefore represented by the circle $X^2+Y^2=1/16$, and spatial infinity corresponds to the unit circle $X^2+Y^2=1$. These coordinates are used only for the graphical representation of the spherical profiles. The color scale and the radial curves represent the ratio between the field and its maximum value,
\[
\frac{f(R)}{f_{\max}},\qquad f_{\max}=\max f(R),
\]
where the maximum is taken over the exterior region, separately for each plotted quantity.

\begin{figure}[!htbp]
    \centering
    \begin{subfigure}[b]{0.36\linewidth}
        \centering
        \includegraphics[width=\linewidth]{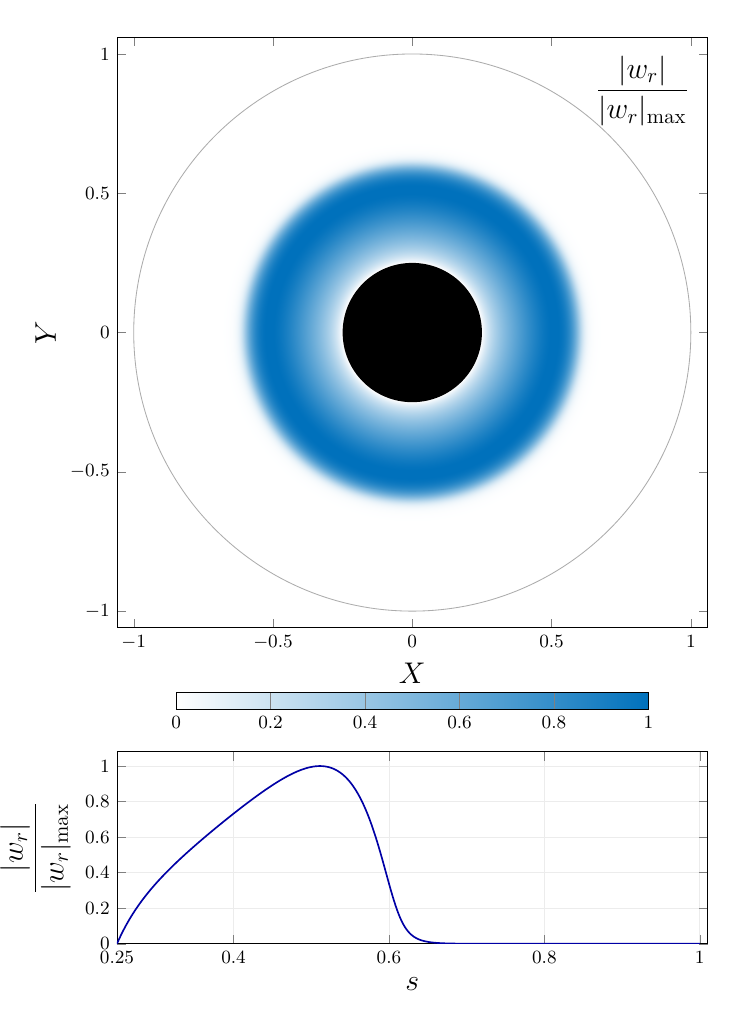}
    \end{subfigure}\hspace{0.08\linewidth}
    \begin{subfigure}[b]{0.36\linewidth}
        \centering
        \includegraphics[width=\linewidth]{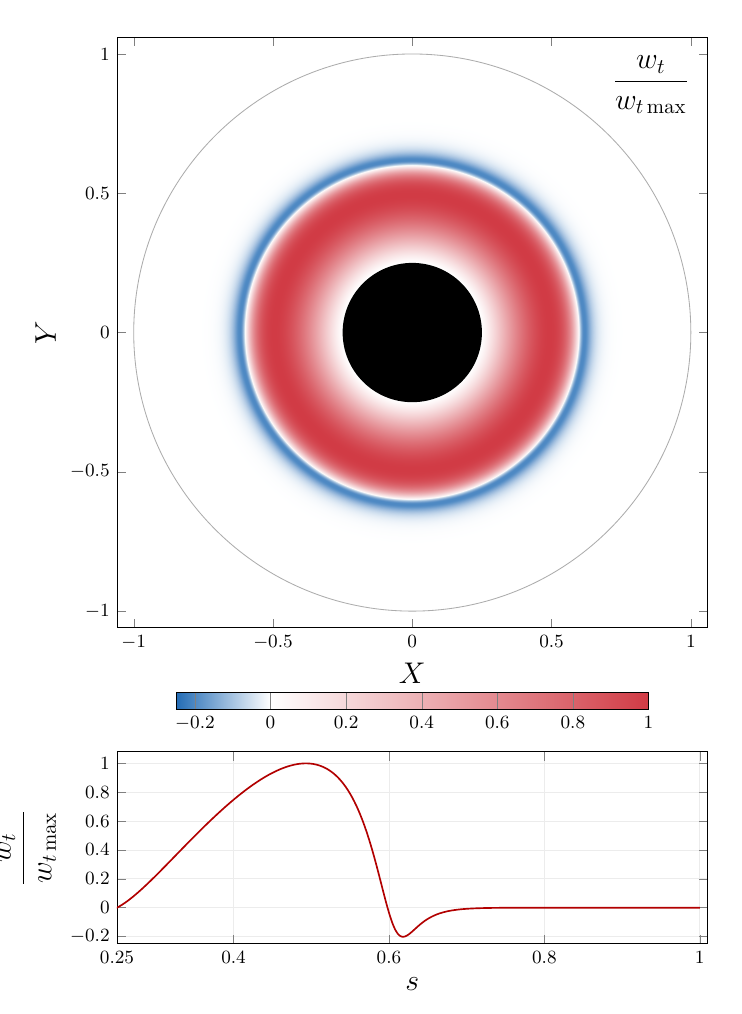}
    \end{subfigure}
    \caption{Normalized profiles of $|w_r|$ (left) and $w_t$ (right) for the solution of Fig.~\ref{fig:radial_profiles}, shown as compactified maps in the $(X,Y)$ plane (top) and as functions of $s$ (bottom). The black disk is the region inside the horizon and the grey circle corresponds to spatial infinity. Both components peak at $s\simeq0.5$; $w_t$ changes sign at $s\simeq0.60$ and has a shallow negative minimum, $w_t\simeq-0.20\,w_{t,\max}$, at $s\simeq0.62$.}
    \label{fig_2d_1}
\end{figure}
\begin{figure}[!htbp]
    \centering
    \begin{subfigure}[b]{0.36\linewidth}
        \centering
        \includegraphics[width=\linewidth]{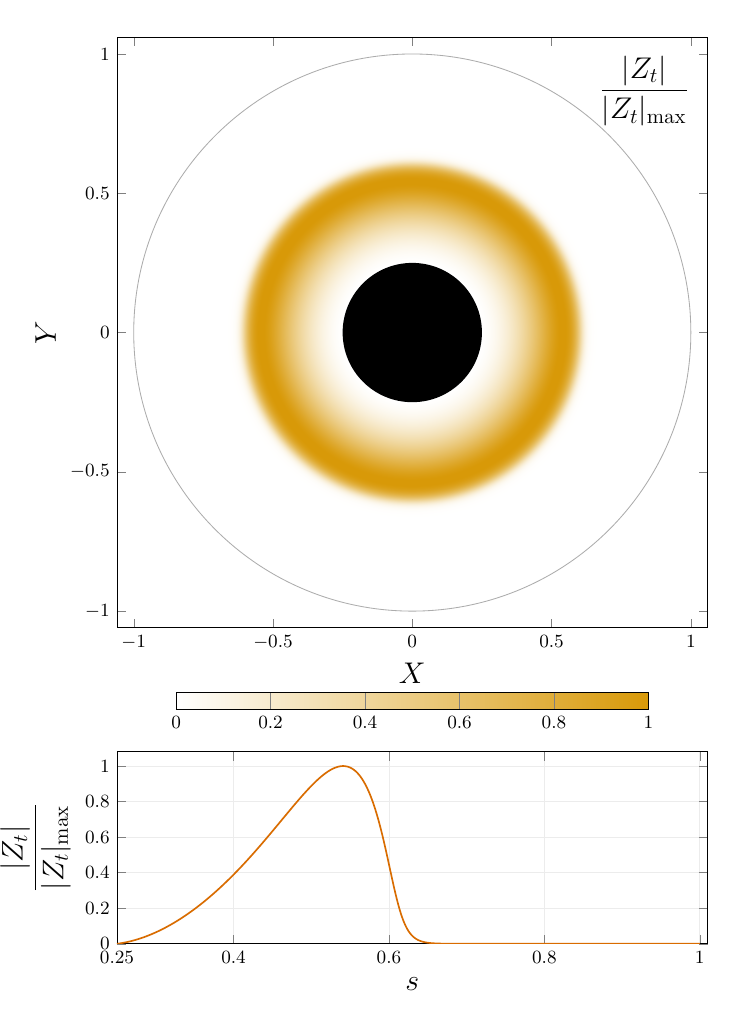}
    \end{subfigure}\hspace{0.08\linewidth}
    \begin{subfigure}[b]{0.36\linewidth}
        \centering
        \includegraphics[width=\linewidth]{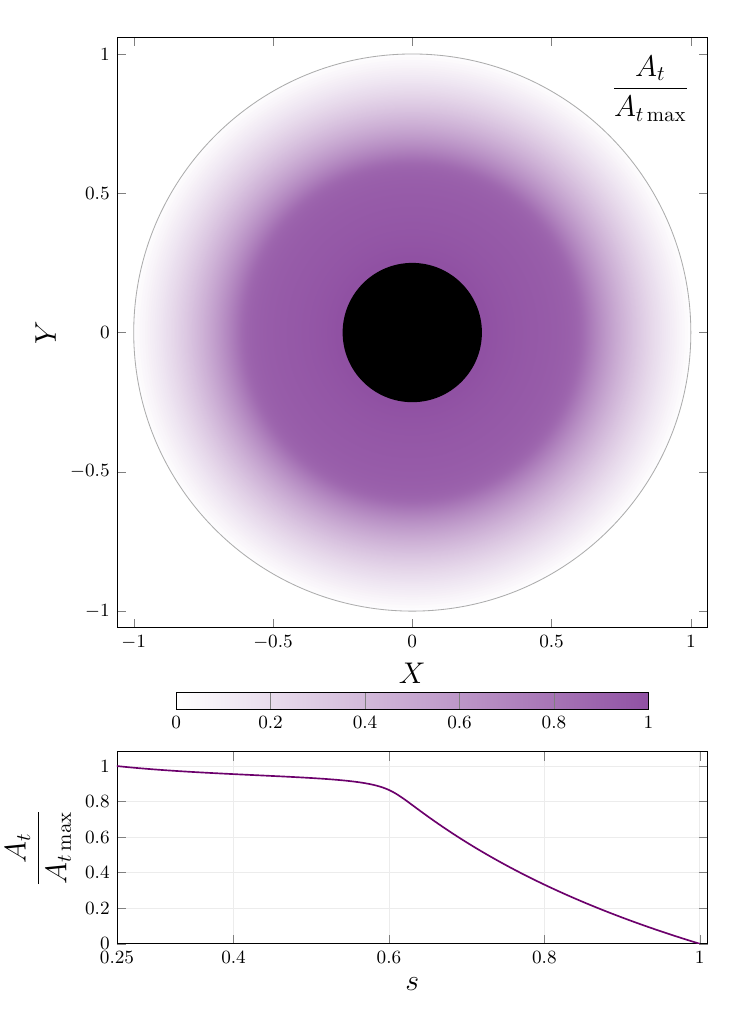}
    \end{subfigure}
    \caption{Same as Fig.~\ref{fig_2d_1} for $|Z_t|$ (left) and $A_t$ (right). The field $Z_t$ is localized in the same shell as the charged vector, peaking at $s\simeq0.54$, whereas $A_t$ is maximal at the horizon, where $A_t=\omega/e$, and decreases monotonically, its Coulomb tail extending up to spatial infinity.}
    \label{fig_2d_2}
\end{figure}

\begin{figure}[!htbp]
    \centering
    \includegraphics[width=0.36\linewidth]{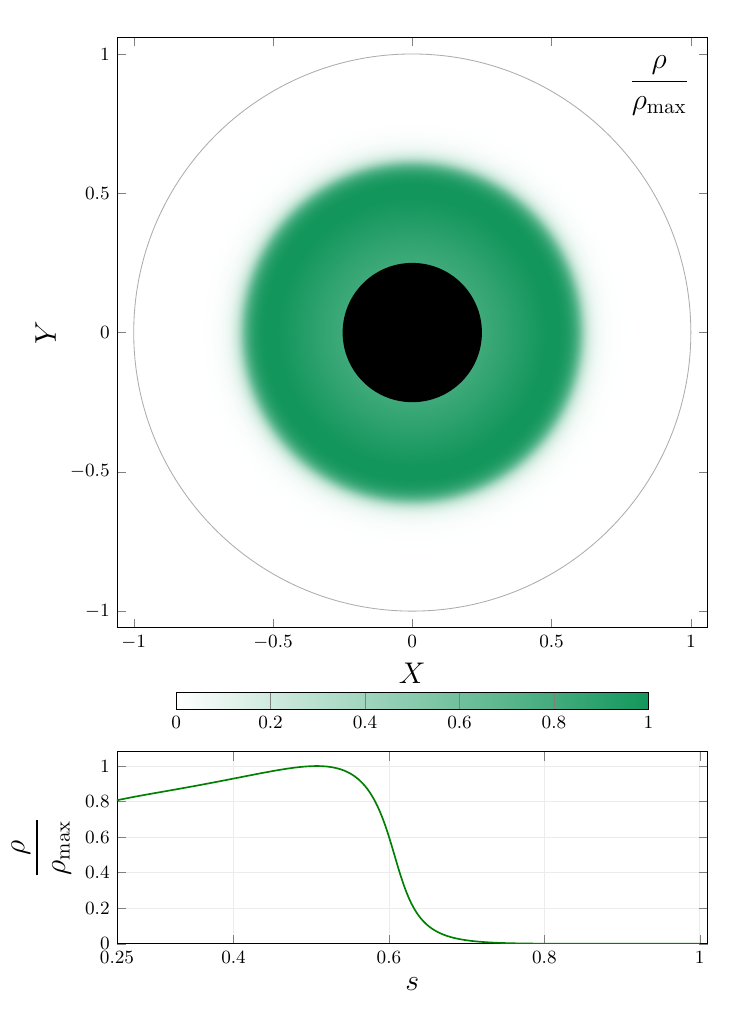}
    \caption{Same as Fig.~\ref{fig_2d_1} for the Higgs vev deviation, $\rho=1-\phi^2$. The Higgs field is suppressed in the same shell: $\rho$ peaks at $s\simeq0.51$, where $\phi\simeq0.72$, equals $0.81\,\rho_{\max}$ at the horizon and becomes negligible for $s\gtrsim0.72$.}
    \label{fig_2d_3}
\end{figure}

For visualization purposes, we define $\rho=1-\phi^2$. The radial component of $w$ and the temporal component of $Z$ are negative. Hence, we plot their absolute values. The sign of the temporal component of $w$ is retained. The resulting maps, for the solution of Fig.~\ref{fig:radial_profiles}, are shown in Figs.~\ref{fig_2d_1}--\ref{fig_2d_3}. All the massive fields are concentrated in a shell around the horizon, with maxima at $0.49\lesssim s\lesssim0.54$, that is, inside the region where $m_{\mathrm{eff}}^2<0$, and decay exponentially after it. The electromagnetic potential, on the other hand, is maximal at the horizon and extends up to infinity, due to its Coulomb tail.

\section{Further remarks}\label{sec:remarks}

In this work, we have studied static, spherically symmetric and asymptotically flat black holes  purely electrically charged in the Einstein-Weinberg-Salam model. These solutions possess a regular event horizon and support nontrivial electroweak fields in the exterior region. Our construction extends the study of the bosonic stars in Ref.~\cite{CostaFilhoGervalle2026} to black hole configurations, with the weak mixing angle fixed to its Standard Model value \cite{PhysRevD.110.030001}.

We have derived the horizon conditions required for the existence of these hairy black holes, in particular the resonance condition. In the gauge adopted for the numerical construction, this condition becomes $\omega=eA_t(r_H)$, relating the frequency of the charged field to the electromagnetic potential at the horizon.  We have also obtained a gauge-invariant mass formula and the corresponding first law.

The equations of motion specialized to this ansatz are displayed in Appendix~\ref{app:eom}. In Appendix~\ref{app:analytic-obstructions}, we have derived integral identities that provide constraints on the solutions whenever they exist. In particular, nontrivial charged-vector hair requires $\omega Q_e>0$ and at least one bounded exterior region in which the effective mass squared defined in Eq.~\eqref{eq:app-effective-mass} becomes negative.

The equations of motion for the spherical ansatz are displayed in Appendix~\ref{app:eom}. The integral identities derived in Appendix~\ref{app:analytic-obstructions} impose necessary conditions for the existence of nontrivial charged-vector hair. We have shown that the frequency and electric charge must have the same sign, $\omega Q_e>0$, and the effective mass squared defined in Eq.~\eqref{eq:app-effective-mass} must become negative in at least one bounded region outside the horizon.

In Appendix~\ref{app:rn-zero-mode}, we have shown that a nonextremal, purely electrically charged RN black hole does not support a regular, exponentially localized, spherically symmetric charged-vector cloud satisfying the resonance condition, at least at the perturbative level. Consequently, within the sector and regularity assumptions considered here, the hairy black holes cannot emerge through a bifurcation from the nonextremal RN family. This distinguishes the present electric sector from the magnetic electroweak branches~\cite{Gervalle:2024yxj,Gervalle2025}.

Within the explored parameter space, we have found two apparently disconnected families of bosonic stars. The first admits the usual dilute Newtonian limit, whereas no weak-field limit was found for the second family. The latter remains strongly gravitating throughout the investigated domain, and all its computed configurations possess a pair of light rings. The hairy black holes constructed here are connected to this second family, which provides part of the boundary of their domain of existence through the limit $r_H\to0$.

The structure of this domain can be followed along sequences of constant frequency. For frequencies above the critical value $\omega_c$ defined in Eq.~\eqref{eq:critical-frequency}, we find sequences connecting two stars with the same frequency: the horizon radius initially increases from zero, reaches a maximum and subsequently decreases to zero again. Below $\omega_c$, the explored sequences do not return to another stellar branch. Instead, the Hawking temperature decreases and suggests an approach to an extremal configuration. Increasing numerical errors prevented further continuation and, therefore, a definite identification of the limiting solutions.

All the black holes obtained numerically violate the RN bound, with $1<\sqrt{\kappa/2}Q_e/M\lesssim1.006$. Thus, none of these configurations shares its mass and electric charge with an RN black hole. Their reduced horizon areas are small, $A_H/(16\pi M^2)\leq0.058$. 
We have also found pairs consisting of a star and a black hole with the same mass and electric charge, but different frequencies. Hence, the global charges $(M,Q_e)$ do not uniquely specify a configuration within the combined set of stellar and black hole solutions. 

Finally, our previous study of bosonic stars \cite{CostaFilhoGervalle2026} and the present investigation of their black hole generalizations were motivated by the search for exotic compact objects within a fundamental theory whose field content is known to exist in Nature. A next step is, therefore, to investigate whether these configurations can reproduce some of the observable properties of astrophysical black holes and access their possible formation mechanisms and stability.

\subsection*{Acknowledgements}
The author is supported by CNPq under the Postdoctoral Junior (PDJ) program, grant 153723/2025-4. This work was partially supported by FCT I.P. under Project 2025.09655.CPCA.A2 through FCCN at Deucalion supercomputer, jointly funded by EuroHPC JU and Portugal.

\clearpage
\appendix

\section{Equations of motion}
\label{app:eom}

In this appendix, we present the equations of motion for the static,
spherically symmetric metric~\eqref{eq:BH-metric-ansatz} and the matter
ansatz~\eqref{eq:BH-matter-ansatz}, whose charged-vector field has harmonic
time dependence, in the gauge specified in
Sec.~\ref{sec:spherical-ansatz}. Throughout this section, a prime denotes
differentiation with respect to $r$. For convenience, we introduce the combinations
\begin{equation}
  \mathcal{E}_A\equiv A_t'-e w_t w_r,\qquad
  \mathcal{E}_Z\equiv Z_t'+g^2 w_t w_r,\qquad
  \mathcal{E}_W\equiv
  w_t'-\left(\omega-eA_t+g^2Z_t\right)w_r,
 \label{eq:app-electric-combinations}
\end{equation}
and write $\mathcal{G}=r^2e^{F_0+3F_1}$. The temporal components of the vector equations~\eqref{eq:maxwell-equation}--\eqref{eq:w-equation} reduce to
\begin{align}
 \left[\mathcal{G}e^{-2F_0-2F_1}\mathcal{E}_A\right]'
 &=e\mathcal{G}e^{-2F_0-2F_1}\mathcal{E}_W w_r,
 \label{eq:app-At-equation}\\
 \left[\mathcal{G}e^{-2F_0-2F_1}\mathcal{E}_Z\right]'
 &=-g^2\mathcal{G}e^{-2F_0-2F_1}\mathcal{E}_W w_r
 +\frac{\mathcal{G}m_{\mathrm{Z}}^2\phi^2e^{-2F_0}}{N}Z_t,
 \label{eq:app-Zt-equation}\\
 \left[\mathcal{G}e^{-2F_0-2F_1}\mathcal{E}_W\right]'
 &=\mathcal{G}e^{-2F_0-2F_1}
 \left(-e\mathcal{E}_A+g^2\mathcal{E}_Z\right)w_r+ \frac{\mathcal{G}m_{\mathrm{W}}^2\phi^2e^{-2F_0}}{N}w_t.
 \label{eq:app-wt-equation}
\end{align}

In turn, the radial component of Eq.~\eqref{eq:w-equation} gives
\begin{equation}
 \left(-e\mathcal{E}_A+g^2\mathcal{E}_Z\right)w_t
 -\left(\omega-eA_t+g^2Z_t\right)\mathcal{E}_W
 -m_{\mathrm{W}}^2\phi^2Ne^{2F_0}w_r=0.
 \label{eq:limit-wr}
\end{equation}

The Higgs equation~\eqref{eq:Higgs-equation} becomes
\begin{equation}
  \left[\mathcal{G}Ne^{-2F_1}\phi'\right]'
  =\frac{\mathcal{G}\phi}{4}
  \left[
    \beta(\phi^2-1)
    +g^2Ne^{-2F_1}w_r^2
    -\frac{e^{-2F_0}}{N}\left(g^2w_t^2+Z_t^2\right)
  \right].
  \label{eq:limit-phi}
\end{equation}

Finally, the vector constraint~\eqref{eq:w-constraint} takes the form
\begin{equation}
  \left[\mathcal{G}Ne^{-2F_1}\phi^2w_r\right]'
  =-\frac{\mathcal{G}e^{-2F_0}}{N}\,
  \phi^2\left(\omega-eA_t-g'^2Z_t\right)w_t.
  \label{eq:limit-constraint}
\end{equation}

This relation follows from the vector equations and does not provide
an additional independent equation. For the metric~\eqref{eq:BH-metric-ansatz}, the nontrivial diagonal
components of Einstein's equations~\eqref{eq:Einstein-equations} are the components $(t,t)$, $(r,r)$ and $(\theta,\theta)$. Explicitly, these read
\begin{align}
 e^{-2F_1}\left[
 N\left(2F_1''+\frac{4F_1'}{r}+F_1'^2\right)
 +N'F_1'\right]
 &=\kappa T^t{}_t,
 \label{eq:limit-Einstein-t}\\
 e^{-2F_1}\left\{
 N\left[2F_0'\left(\frac{1}{r}+F_1'\right)
 +\frac{2F_1'}{r}+F_1'^2\right]+N'F_1'\right\}
 &=\kappa T^r{}_r,
 \label{eq:limit-Einstein-r}\\
 e^{-2F_1}\left\{
 N\left[F_0''+F_1''+F_0'^2
 +\frac{F_0'+F_1'}{r}\right]
 +\frac{N'}{2}\left(3F_0'+F_1'\right)\right\}
 &=\kappa T^\theta{}_\theta.
 \label{eq:limit-Einstein-angular}
\end{align}

The corresponding components of the energy-momentum tensor are
\begin{align}
- T^t{}_t={}&
 \frac{e^{-2F_0-2F_1}}{2}
 \left(\mathcal{E}_A^2+\mathcal{E}_Z^2+\mathcal{E}_W^2\right)
 +Ne^{-2F_1}\phi'^2
 +\frac{\beta}{8}\left(\phi^2-1\right)^2
 \notag\\
 &+\frac{g^2Ne^{-2F_1}}{4}\phi^2w_r^2
 +\frac{e^{-2F_0}}{4N}\phi^2
 \left(g^2w_t^2+Z_t^2\right),
 \label{eq:limit-rho}\\
 T^r{}_r-T^t{}_t={}&
 2Ne^{-2F_1}\phi'^2
 +\frac{g^2Ne^{-2F_1}}{2}\phi^2w_r^2
 +\frac{e^{-2F_0}}{2N}\phi^2
 \left(g^2w_t^2+Z_t^2\right),
 \label{eq:limit-rho-p}\\
 T^\theta{}_\theta={}&
 \frac{e^{-2F_0-2F_1}}{2}
 \left(\mathcal{E}_A^2+\mathcal{E}_Z^2+\mathcal{E}_W^2\right)
 -Ne^{-2F_1}\phi'^2
 -\frac{\beta}{8}\left(\phi^2-1\right)^2
 \notag\\
 &-\frac{g^2Ne^{-2F_1}}{4}\phi^2w_r^2
 +\frac{e^{-2F_0}}{4N}\phi^2
 \left(g^2w_t^2+Z_t^2\right).
 \label{eq:limit-angular-stress}
\end{align}
In particular, subtracting Eq.~\eqref{eq:limit-Einstein-t} from
Eq.~\eqref{eq:limit-Einstein-r} yields 
\begin{equation}
 2Ne^{-2F_1}\left[
 \frac{F_0'-F_1'}{r}+F_0'F_1'-F_1''\right]
 =\kappa\left(T^r{}_r-T^t{}_t\right).
 \label{eq:app-Einstein-difference}
\end{equation}

Equations~\eqref{eq:app-At-equation}--\eqref{eq:limit-Einstein-angular} form a system of nine equations for the seven radial functions~\eqref{eq:BH-radial-functions}. Numerically, we solve a combination of these equations forming a system of seven equations, while the remaining two are used as constraints. That is, they are not solved directly but are used as a consistency check on the solutions.

\section{Integral identities}
\label{app:analytic-obstructions}

This appendix derives two integral identities that constrain the spherical electric sector. The first fixes the relative sign of the frequency and electric charge, while the second establishes that nontrivial charged-vector hair requires a region of negative effective mass squared outside the horizon. Throughout, we assume a nonextremal black hole, a regular unitary-gauge exterior with $\phi>0$, and metric and matter functions analytic near the horizon. We comment that similar integral identities also appear in the context of no-go theorems \cite{GaltsovErshov1989,Bizon:1992pi,Greene:1992fw,HerdeiroRadu2020}.


\subsection{A positive identity and the sign of the charge}
\label{app:positive-identity}

Multiply Eqs.~\eqref{eq:app-At-equation},
\eqref{eq:app-Zt-equation}, and \eqref{eq:app-wt-equation} by
$A_t-\omega/e$, $Z_t$, and $w_t$, respectively, and add them.  Since
\begin{equation}
 e\left(A_t-\frac{\omega}{e}\right)-g^2Z_t=-\left(\omega-eA_t+g^2Z_t\right),
 \label{eq:app-frequency-rearrangement}
\end{equation}
the product rule gives
\begin{align}
 \frac{\mathrm{d}}{\mathrm{d} r}\Bigg\{
 \mathcal{G} e^{-2F_0-2F_1}\Bigg[&
 \left(A_t-\frac{\omega}{e}\right)\mathcal{E}_A
 +Z_t\mathcal{E}_Z+w_t\mathcal{E}_W\Bigg]\Bigg\}
 \notag\\
 =\mathcal{G} e^{-2F_0-2F_1}\Big[&
 A_t'\mathcal{E}_A+Z_t'\mathcal{E}_Z+w_t'\mathcal{E}_W
 -\left(\omega-eA_t+g^2Z_t\right)w_r\mathcal{E}_W
 \notag\\
 &+\left(-e\mathcal{E}_A+g^2\mathcal{E}_Z\right)w_tw_r\Big]
 +\frac{\mathcal{G}\phi^2e^{-2F_0}}{N}
 \left(m_{\mathrm{W}}^2w_t^2+m_{\mathrm{Z}}^2Z_t^2\right).
 \label{eq:app-positive-product-rule}
\end{align}

The definitions \eqref{eq:app-electric-combinations} imply
\begin{align}
 A_t'\mathcal{E}_A+Z_t'\mathcal{E}_Z+w_t'\mathcal{E}_W
 ={}&\mathcal{E}_A^2+\mathcal{E}_Z^2+\mathcal{E}_W^2
 \notag\\
 &+w_tw_r\left(e\mathcal{E}_A-g^2\mathcal{E}_Z\right)
 +\left(\omega-eA_t+g^2Z_t\right)w_r\mathcal{E}_W.
 \label{eq:app-product-rule-expansion}
\end{align}

Substitution into Eq.~\eqref{eq:app-positive-product-rule} cancels all mixed
terms and yields
\begin{align}
 \frac{\mathrm{d}}{\mathrm{d} r}\Bigg\{
 \mathcal{G} e^{-2F_0-2F_1}\Bigg[&
 \left(A_t-\frac{\omega}{e}\right)\mathcal{E}_A
 +Z_t\mathcal{E}_Z+w_t\mathcal{E}_W\Bigg]\Bigg\}
 \notag\\
 ={}&\mathcal{G} e^{-2F_0-2F_1}\left(
 \mathcal{E}_A^2+\mathcal{E}_Z^2+\mathcal{E}_W^2\right)+\frac{\mathcal{G}\phi^2e^{-2F_0}}{N}
 \left(m_{\mathrm{W}}^2w_t^2+m_{\mathrm{Z}}^2Z_t^2\right).
 \label{eq:app-local-positive-identity}
\end{align}

Integrating Eq.~\eqref{eq:app-local-positive-identity} from the horizon
to infinity and using the horizon boundary conditions together with the
asymptotic behavior specified in Sec.~\ref{sec:spherical-ansatz},
we obtain
\begin{align}
 \frac{\omega Q_e}{e}
 =\int_{r_H}^{\infty}\mathrm{d} r\,\mathcal{G}\Bigg[&
 e^{-2F_0-2F_1}\left(
 \mathcal{E}_A^2+\mathcal{E}_Z^2+\mathcal{E}_W^2\right)+\frac{\phi^2e^{-2F_0}}{N}
 \left(m_{\mathrm{W}}^2w_t^2+m_{\mathrm{Z}}^2Z_t^2\right)\Bigg]\geq0.
 \label{eq:app-hairy-positive-identity}
\end{align}

 Since each term in the integrand is nonnegative,  taking $e>0$, every solution with nontrivial  hair  obeys
\begin{equation}
 \omega Q_e>0.
 \label{eq:app-frequency-charge-sign}
\end{equation}

Thus, changing the sign of the frequency $\omega$ also changes the sign of the electric charge, with $\omega>0$ and $\omega<0$ corresponding to positively and negatively charged solutions, respectively. Finally, nontrivial solutions require both the frequency $\omega$ and the electric charge $Q_e$ to be nonzero.

\subsection{A necessary effective mass}
\label{app:effective}

Multiply Eq.~\eqref{eq:app-wt-equation} by $w_t$ and integrate by parts. Using that the fields decay exponentially at infinity and the boundary conditions
 \eqref{eq:basic-matter-horizon-BC} and \eqref{eq:horizon-evenness-conditions}, we find
\begin{align}
 0=\int_{r_H}^{\infty}\mathrm{d} r\,\mathcal{G}\Bigg[
 e^{-2F_0-2F_1}w_t'\mathcal{E}_W
 +e^{-2F_0-2F_1}
 \left(-e\mathcal{E}_A+g^2\mathcal{E}_Z\right)w_tw_r +\frac{m_{\mathrm{W}}^2\phi^2e^{-2F_0}}{N}w_t^2\Bigg].
 \label{eq:app-integrated-wt-equation}
\end{align}

Multiplying Eq.~\eqref{eq:limit-wr} by $w_r$ gives
\begin{align}
 e^{-2F_0-2F_1}
 \left(-e\mathcal{E}_A+g^2\mathcal{E}_Z\right)w_tw_r
 ={}e^{-2F_0-2F_1}\left(\omega-eA_t+g^2Z_t\right)w_r\mathcal{E}_W +m_{\mathrm{W}}^2\phi^2Ne^{-2F_1}w_r^2.
 \label{eq:app-wr-substitution}
\end{align}

Substitute this into Eq.~\eqref{eq:app-integrated-wt-equation} and use
\begin{equation}
 \left(w_t'+\left(\omega-eA_t+g^2Z_t\right)w_r\right)\mathcal{E}_W
 =(w_t')^2-\left(\omega-eA_t+g^2Z_t\right)^2w_r^2.
 \label{eq:app-W-elementary-identity}
\end{equation}

One obtains
\begin{align}
 0=\int_{r_H}^{\infty}\mathrm{d} r\,\mathcal{G}\Bigg[&
 e^{-2F_0-2F_1}(w_t')^2
 +\frac{m_{\mathrm{W}}^2\phi^2e^{-2F_0}}{N}w_t^2
 \notag\\[-1mm]
 &+\left(m_{\mathrm{W}}^2\phi^2Ne^{-2F_1}
 -e^{-2F_0-2F_1}\left(\omega-eA_t+g^2Z_t\right)^2\right)w_r^2\Bigg].
 \label{eq:app-W-integral-identity}
\end{align}

Hence, every solution with nontrivial  hair must have at least one exterior point at which
\begin{equation}
 \frac{\left[\omega-eA_t(r)+g^2Z_t(r)\right]^2}{N e^{2F_0}}
 >m_{\mathrm{W}}^2\phi^2(r).
 \label{eq:app-necessary}
\end{equation}

Equivalently, define the diagnostic radial function
\begin{equation}
 m_{\mathrm{eff}}^2(r)\equiv m_{\mathrm{W}}^2\phi^2(r)
 -\frac{\left[\omega-eA_t(r)+g^2Z_t(r)\right]^2}{N e^{2F_0}}.
 \label{eq:app-effective-mass}
\end{equation}

Thus, for a solution with nontrivial charged-vector hair, there must be one
or more bounded intermediate regions in which $m_{\mathrm{eff}}^2<0$.

\section{Reissner-Nordstr\"om clouds?}
\label{app:rn-zero-mode}

This appendix investigates whether a branch of spherically symmetric, purely electric hairy black holes can bifurcate perturbatively from the nonextremal Reissner-Nordström family. We follow the integral-identity argument of Appendix \ref{app:effective}, now applied to the linearized charged-vector equations on a fixed RN background. We use an areal radius $R$, distinct from the coordinate $r$ of Eq.~\eqref{eq:BH-metric-ansatz}. In the conventions of this paper, the RN background reads
\begin{equation}
 \begin{aligned}
 \mathrm{d}s^2&=-f(R)\,\mathrm{d}t^2
 +\frac{\mathrm{d}R^2}{f(R)}+R^2\mathrm{d}\Omega_2^2,\\
 A&=\frac{Q_e}{R}\,\mathrm{d}t,
 \qquad Z=w=0,
 \qquad \phi=1,
 \end{aligned}
 \label{eq:app-RN-background}
\end{equation}
with
\begin{equation}
 f(R)=1-\frac{2M}{R}+\frac{\kappa Q_e^2}{2R^2}
 =\left(1-\frac{R_+}{R}\right)
  \left(1-\frac{R_-}{R}\right),\qquad R_\pm=M\pm\sqrt{M^2-\frac{\kappa Q_e^2}{2}}.
 \label{eq:app-RN-lapse}
\end{equation}

We restrict attention to the nonextremal case, $R_+>R_-$. Then
 \begin{equation}
 M^2>\frac{\kappa Q_e^2}{2},\qquad
 R_++R_-=2M, \qquad
  R_+R_-=\frac{\kappa Q_e^2}{2}.
 \label{eq:app-RN-roots}
\end{equation}

Because the charged vector enters the neutral equations and the
energy-momentum tensor at least quadratically, its linear perturbation decouples from the
perturbations of the metric and of the neutral fields, which are of order
$\epsilon^2$.  A spherical perturbation can be written
\begin{equation}
 w=\epsilon e^{-\mathrm{i}\omega t}
 \left[w_t(R)\,\mathrm{d}t
 +\mathrm{i}w_r(R)\,\mathrm{d}R\right]
 +\mathcal{O}(\epsilon^2),
 \label{eq:app-RN-W-perturbation}
\end{equation}
where $w_t$ and $w_r$ are real radial profiles. To first order,
Eq.~\eqref{eq:w-equation} and the Lorenz-like condition~\eqref{eq:w-constraint} reduce to
\begin{align}
 \mathcal{D}_\mu w^{\mu\nu}+\mathrm{i}eF^{\nu\sigma}w_\sigma
 &=m_{\mathrm{W}}^2w^\nu,
 \label{eq:app-linear-W-equation}\\
 \mathcal{D}_\mu w^\mu&=0.
 \label{eq:app-linear-W-constraint}
\end{align}

The temporal and radial components of Eq.~\eqref{eq:app-linear-W-equation} are
\begin{align}
 \frac{1}{R^2}\left[R^2\left(w_t'
 -\left(\omega-\frac{eQ_e}{R}\right)w_r\right)\right]'
 -\frac{eQ_e}{R^2}w_r
 &=\frac{m_{\mathrm{W}}^2}{f}w_t,
 \label{eq:app-RN-time-equation}\\
 -\left(\omega-\frac{eQ_e}{R}\right)
 \left(w_t'-\left(\omega-\frac{eQ_e}{R}\right)w_r\right)
 +\frac{eQ_e}{R^2}w_t
 &=m_{\mathrm{W}}^2fw_r,
 \label{eq:app-RN-radial-equation}
\end{align}
while the Lorenz-like constraint becomes
\begin{equation}
 \frac{1}{R^2}\left(R^2fw_r\right)'
 +\frac{\omega-eQ_e/R}{f}w_t=0.
 \label{eq:app-RN-Lorenz-constraint}
\end{equation}

For a candidate mode satisfying the horizon conditions of
Sec.~\ref{sec:covariant-horizon}, the boundary
conditions~\eqref{eq:basic-matter-horizon-BC} apply to the leading-order fields
at the horizon, $R=R_+$.  The resonance condition $A_t(R_+)=\omega/e$ fixes
\begin{equation}
 \omega=\frac{eQ_e}{R_+}.
 \label{eq:app-RN-synchronization}
\end{equation}

The remaining horizon condition is $w_t(R_+)=0$. We assume regular
near-horizon series expansions and set $y=R-R_+$. By
Eq.~\eqref{eq:app-RN-synchronization},
$\omega-eQ_e/R=\omega y/R=\mathcal{O}(y)$, so the second term in
Eq.~\eqref{eq:app-RN-Lorenz-constraint} is $\mathcal{O}(y)$. At the
horizon, the constraint therefore reduces to
$f'(R_+)w_r(R_+)=0$. Since
$f'(R_+)=(R_+-R_-)/R_+^2>0$, also $w_r(R_+)=0$. Hence,
$w_t=w_{t,1}y+\mathcal{O}(y^2)$ and
$w_r=w_{r,1}y+\mathcal{O}(y^2)$, so that
$w_t'-(\omega-eQ_e/R)w_r=w_{t,1}+\mathcal{O}(y)$ remains finite and
$R^2w_t[w_t'-(\omega-eQ_e/R)w_r]=\mathcal{O}(y)$.

We multiply Eq.~\eqref{eq:app-RN-time-equation} by $R^2w_t$ and
integrate from $R_+$ to infinity. Exponential decay removes the boundary
term at infinity, while the near-horizon behavior above removes the
horizon term. Integration by parts then gives
\begin{equation}
 0=\int_{R_+}^{\infty}\mathrm{d}R\,R^2\left[
 w_t'\left(w_t'-\left(\omega-\frac{eQ_e}{R}\right)w_r\right)
 +\frac{eQ_e}{R^2}w_tw_r+\frac{m_{\mathrm{W}}^2}{f}w_t^2
 \right].
 \label{eq:app-RN-first-integral}
\end{equation}
Multiplying Eq.~\eqref{eq:app-RN-radial-equation} by $w_r$ and
rearranging gives
\begin{equation}
 -\left(\omega-\frac{eQ_e}{R}\right)w_t'w_r
 +\frac{eQ_e}{R^2}w_tw_r
 =\left[m_{\mathrm{W}}^2f
 -\left(\omega-\frac{eQ_e}{R}\right)^2\right]w_r^2.
 \label{eq:app-RN-cross-term}
\end{equation}
Consequently,
\begin{equation}
 0=\int_{R_+}^{\infty}\mathrm{d}R\,R^2\left[
 (w_t')^2+\frac{m_{\mathrm{W}}^2}{f}w_t^2
 +\left(m_{\mathrm{W}}^2f
 -\left(\omega-\frac{eQ_e}{R}\right)^2\right)w_r^2
 \right].
 \label{eq:app-RN-positive-identity}
\end{equation}

The resonance condition~\eqref{eq:app-RN-synchronization} implies
\begin{equation}
 \omega-\frac{eQ_e}{R}
 =\omega\left(1-\frac{R_+}{R}\right),
 \qquad
 \frac{\left(\omega-\frac{eQ_e}{R}\right)^2}{f}
 =\omega^2\frac{R-R_+}{R-R_-}\leq\omega^2
 \qquad (R>R_+).
 \label{eq:app-RN-frequency-bound}
\end{equation}
Since $0\leq(R-R_+)/(R-R_-)<1$ for $R\geq R_+>R_-$, the inequality
is strict whenever $\omega\neq0$, corresponding to a charged background.
For Schwarzschild, $Q_e=\omega=0$ and both sides vanish.
Exponential localization requires $|\omega|<m_{\mathrm{W}}$. Hence,
\begin{equation}
 m_{\mathrm{W}}^2f-\left(\omega-\frac{eQ_e}{R}\right)^2
 =f\left[m_{\mathrm{W}}^2
 -\frac{\left(\omega-\frac{eQ_e}{R}\right)^2}{f}\right]>0
 \qquad (R>R_+).
 \label{eq:app-RN-positive-coefficient}
\end{equation}
All three terms in Eq.~\eqref{eq:app-RN-positive-identity} are
nonnegative, and the coefficients of $w_t^2$ and $w_r^2$ are strictly
positive in the open exterior. Therefore,
\begin{equation}
 w_t(R)=w_r(R)=0.
 \label{eq:app-RN-trivial-mode}
\end{equation}

Thus, no nontrivial, regular, exponentially localized, spherically symmetric $W$ cloud exists
on a nonextremal RN black hole. Since the charged-vector perturbation decouples at linear order, an ordinary $W$-hairy branch admitting a regular near-horizon series expansion cannot bifurcate perturbatively from RN.  The absence of linear clouds should be distinguished from the existence of nonlinear configurations in the probe approximation. Indeed, nonlinear electroweak clouds have been constructed on Schwarzschild backgrounds \cite{HerdeiroRaduEWClouds2026}.

\bibliographystyle{hhieeetr}
\bibliography{biblio}

\end{document}